\documentclass[fleqn,usenatbib]{mnras}

\usepackage{newtxtext,newtxmath}

\usepackage[T1]{fontenc}
\usepackage{ae,aecompl}

\usepackage{graphicx}	
\usepackage{amsmath}	
\usepackage{amssymb}	
\usepackage{multicol}        
\usepackage{bm}		
\usepackage{pdflscape}	
\usepackage{soul}
\usepackage{enumitem}
\usepackage{pdflscape}	
\usepackage[dvipsnames]{xcolor}

\newcommand{\dd}{\mathrm{d}}
\newcommand{\Mpc}{\mathrm{Mpc}}
\newcommand{\Myr}{\mathrm{Myr}}

\newcommand{\kpc}{\mathrm{kpc}}
\newcommand{\pc}{\mathrm{pc}}

\newcommand{\hMpc}{h^{-1} \mathrm{\Mpc}}
\newcommand{\Msol}{\textup{M}_\mathrm{\sun}}

\newcommand{\xMsol}[2]{\ensuremath{{#1}\times 10^{#2} \,\Msol}}
\newcommand{\xScientific}[2]{\ensuremath{{#1} \times 10^{#2}}}

\newcommand{\G}{\mathrm{G}}
\newcommand{\baryonfraction}{f_{\mathrm{b}}}
\newcommand{\fmb}{f_{\mathrm{mb}}}
\newcommand{\mdm}{m_{\mathrm{DM}}}
\newcommand{\softening}{\Delta x}

\newcommand{\zform}{z_{\text{form}}}
\newcommand{\Mvir}{M_{200}}

\newcommand{\tdyn}{\tau_{\text{dyn}}}
\newcommand{\Mgrowth}{\big \langle \frac{ \dd \Mvir}{\dd t} \big \rangle_{\text{dyn}}}
\newcommand{\Mgrowthinline}{\langle \dd \Mvir / \dd t \rangle_{\text{dyn}}}
\newcommand{\SFR}{\mathrm{SFR}}
\newcommand{\efficiency}{\mathit{\epsilon}}
\newcommand{\effnorm}{\mathit{\epsilon}_{\mathrm{N}}}
\newcommand{\effzero}{\mathit{\epsilon}_{\mathrm{0}}}
\newcommand{\effz}{\mathit{\epsilon}_{\mathrm{z}}}

\newcommand{\lmslope}{\mathit{\beta}}
\newcommand{\lmslopezero}{\mathit{\beta}_{\mathrm{0}}}
\newcommand{\lmslopez}{\mathit{\beta}_{\mathrm{z}}}

\newcommand{\hmslope}{\mathit{\gamma}}
\newcommand{\hmslopezero}{\mathit{\gamma}_{\mathrm{0}}}
\newcommand{\hmslopez}{\mathit{\gamma}_{\mathrm{z}}}

\newcommand{\Mone}{M_{\mathrm{1}}}
\newcommand{\Mzero}{M_{\mathrm{0}}}
\newcommand{\Mz}{M_{\mathrm{z}}}

\newcommand{\Mstar}{M_{\star}}
\newcommand{\rvir}{r_{200}}

\newcommand{\lateformer}{\textit{Late-former} }
\newcommand{\mwlike}{\textit{Milky Way-like} }

\newcommand{\MR}[1]{{\color{black}{#1}}}

\title[How cosmological histories shape stellar haloes]
{How cosmological merger histories shape the diversity of stellar haloes}

\author[M. P. Rey \& T. K. Starkenburg]
{Martin P. Rey,$^{1, 2}$\thanks{E-mail: \href{martin.rey@physics.ox.ac.uk}{martin.rey@physics.ox.ac.uk}} and Tjitske K. Starkenburg$^3$
\vspace{0.8mm}
\\
$^{1}$Lund Observatory, Department of Astronomy and Theoretical Physics, Lund University, Box 43, SE-221 00, Lund, Sweden \\
$^{2}$ Sub-department of Astrophysics, University of Oxford, DWB, Keble Road, Oxford OX1 3RH, UK \\ 
$^3$Center for Interdisciplinary Exploration and Research in Astrophysics (CIERA) and \\ Department of Physics and Astronomy,
Northwestern University, 1800 Sherman Ave, Evanston IL 60201, USA
}

\date{Submitted to MNRAS}

\pubyear{2021}

\begin{document}
\label{firstpage}
\pagerange{\pageref{firstpage}--\pageref{lastpage}}
\maketitle

\begin{abstract}
We introduce and apply a new approach to probe the response of galactic stellar haloes to the interplay between cosmological merger histories and galaxy formation physics. We perform dark matter-only, zoomed simulations of two Milky Way-mass hosts and make targeted, controlled changes to their cosmological histories using the genetic modification technique. Populating each history's stellar halo with a semi-empirical, particle tagging approach then enables a controlled study, with all instances converging to the same large-scale structure, dynamical and stellar mass at $z=0$ as their reference. These related merger scenarios alone generate an extended spread in stellar halo mass fractions (1.5 dex) comparable to the observed population, with the largest scatter achieved by growing late ($z\leq1$) major mergers that spread out existing stars to create massive, \textit{in-situ} dominated stellar haloes. Increasing a last major merger at $z\sim2$ brings more accreted stars into the inner regions, resulting in smaller scatter in the outskirts which are predominantly built by subsequent minor events. Exploiting the flexibility of our semi-empirical approach, we show that the diversity of stellar halo masses across scenarios is reduced by allowing shallower slopes in the stellar mass--halo mass relation for dwarf galaxies, while it remains conserved when central stars are born with hotter kinematics across cosmic time. The merger-dependent diversity of stellar haloes thus responds distinctly to assumptions in modelling the central and dwarf galaxies respectively, opening exciting prospects to constrain star formation and feedback at different galactic mass-scales with the coming generation of deep, photometric observatories.
\end{abstract}




\section{Introduction}

Diffuse stellar haloes surrounding galaxies provide a unique window into their cosmological formation and evolution. They assemble over time as a natural consequence of hierarchical $\Lambda$CDM structure formation, originating from both the stripping of stars within external, accreted galaxies (e.g. \citealt{Johnston1996, Johnston1998, Helmi1999, Bullock2005, Read2006StellarHalos, Cooper2010}) and the heating of central, internal stars onto wide orbits (e.g. \citealt{Zolotov2009, Font2011, Cooper2015}). Long dynamical timescales for stars in galactic outskirts preserve memory of these past events (\citealt{Bullock2005}), providing us with a promising path to unravel galactic histories while constraining how star formation is regulated across the spectrum of stellar halo progenitors.

Such archaeological analyses are particularly mature in the Milky Way, where signatures of past accretion events can be pinpointed using overlapping information between complementary datasets in phase space, chemical abundances and ages (e.g. \citealt{Ibata1994Sgt, Helmi1999, Helmi1999Streams, Belokurov2006FieldOfStreams, Belokurov2018, Helmi2018, Bonaca2021,Naidu2021GSECharacterisation}). This wealth of information allows us to narrow down the specific history of our Galaxy (see \citealt{Bland-Hawthorn2016,Helmi2020} for reviews), but it remains challenging to turn it into constraints on the formation and evolution of the general galaxy population, for which larger, extragalactic, samples are required.

Meeting this promise however requires predictions capable to sample many, diverse cosmological merger histories, while resolving astrophysical processes within the interstellar medium of both the central galaxy and dwarf progenitors to the stellar halo. These two requirements place strong, conflicting demands on the best-use of computing time, forcing us to trade between number of objects and physical fidelity to make calculations tractable (see \citealt{Somerville2015} for a review).

Complementary approaches have thus been developed to quantify the diversity of hierarchical stellar haloes. One can focus on sampling many histories and environments to establish correlations between dark matter and stellar halo assemblies, favouring stronger statistics with fast approximations relying on gravity-only simulations (e.g. \citealt{Cooper2013}), or following galaxy formation through more expensive, large hydrodynamical simulations (e.g. \citealt{Pillepich2014, Rodriguez-Gomez2016, Elias2018, Davison2020, Merritt2020, Remus2021}). Alternatively, numerical effort can be concentrated into a handful of objects using zoomed simulations, thus affording more detailed modelling of the interstellar medium while retaining self-consistent cosmological merger histories (e.g. \citealt{Zolotov2009, Font2011, Sanderson2018, Monachesi2019, Font2020, Keller2021}).

In both approaches, however, each galaxy's history and environment are seeded from random, inflationary perturbations in the early-Universe. By construction, every degree of freedom is modified from one object to the next, making it challenging to establish causal relationships between a given merger history and stellar halo observables. This difficulty couples with the sensitivity of stellar haloes to uncertain interstellar medium physics (e.g. \citealt{Font2020, Keller2021}, see \citealt{Naab2017} for a review), further complicating interpretations. In this work, we develop an adaptive technique to tackle this challenge, aiming to vary cosmological merger histories and galaxy formation physics independently, to establish a causal account of how their interplay affects stellar haloes. 

To this end, we perform zoomed, dark matter-only cosmological simulations which we complement with the genetic modification technique (\citealt{Roth2016, Rey2018, Stopyra2021}). Genetic modifications allow us to create different versions of a chosen galaxy, each version with carefully specified modifications to its merger history while maintaining similar cosmological large-scale structure (\citealt{Roth2016, Rey2019VarianceDMOs}). Comparing between scenarios then enables a causal account of how merger histories affect a galaxy's observables (e.g. \citealt{Pontzen2017, Rey2019UFDScatter}), in our case the stellar halo. 

To allow wide exploration of genetically modified scenarios \MR{and physical models}, we populate the stars and stellar halo of each dark matter-only simulation with an efficient `particle tagging' approach. Particle tagging constructs stellar haloes assuming that the phase-space evolution of stars over time can be tracked by a carefully chosen set of dark matter particles extracted from the gravity-only simulation (e.g. \citealt{Bullock2005, DeLucia2008, Cooper2010, Libeskind2011, Rashkov2012, Amorisco2017Atlas,Deason2021DwarfHalos}). With this approach, efficiency is traded for  dynamical approximations due to the dark matter-only context. For instance, kinematics of dark matter particles are unable to self-consistently represent the growth of a central rotating disc, and its subsequent impact on the dynamics and stripping of satellites (see further discussions in \citealt{Cooper2010, Bailin2014, Cooper2017}). Despite these approximations, particle tagging techniques that flag a set of particles regularly, i.e. over a star formation history rather than once at infall, can realistically capture the dynamical diffusion of halo stars in phase space and are able to reproduce stellar haloes obtained from hydrodynamical simulations (\citealt{LeBret2017, Cooper2017}). 

In fact, larger uncertainties in the resulting stellar halo are associated with assigning star formation histories to a dark matter-only simulation than with dynamical approximations (\citealt{Cooper2017}). To further enable efficient exploration of this aspect, we make use of recent data-driven, empirical models that associate each dark matter progenitor with an analytical, parametrized star formation history whose functional form is constrained to reproduce the distributions of observed colours, star formation rates, stellar masses and clustering of the galaxy population (e.g. \citealt{Moster2018, Behroozi2019, OLeary2021}). This enables fast, readily constructed star formation histories for individual, zoomed simulations whose parameters can be conveniently varied within data constraints.

We describe this approach in Section~\ref{sec:methods} and how we apply it across a suite of genetically modified Milky Way-mass dark matter hosts (Section~\ref{sec:gms}). We present the response of stellar haloes to our genetically modified scenarios in Section~\ref{sec:stellarhaloes}, showing how late, violent mergers scattering the pre-existing \textit{in-situ} component are key to reproducing the extent of the observed diversity in stellar halo mass fraction. We then demonstrate how the dwarf galaxy stellar mass--halo mass relation (Section~\ref{sec:dwarfs}) and the birth kinematics of stars (Section~\ref{sec:boundfraction}) both modify absolute stellar halo masses but affect the merger-dependent diversity distinctly. We present our conclusions in Section~\ref{sec:conclu}.

\section{Methods} \label{sec:methods}

\begin{figure*}
  \centering
    \includegraphics[width=\textwidth]{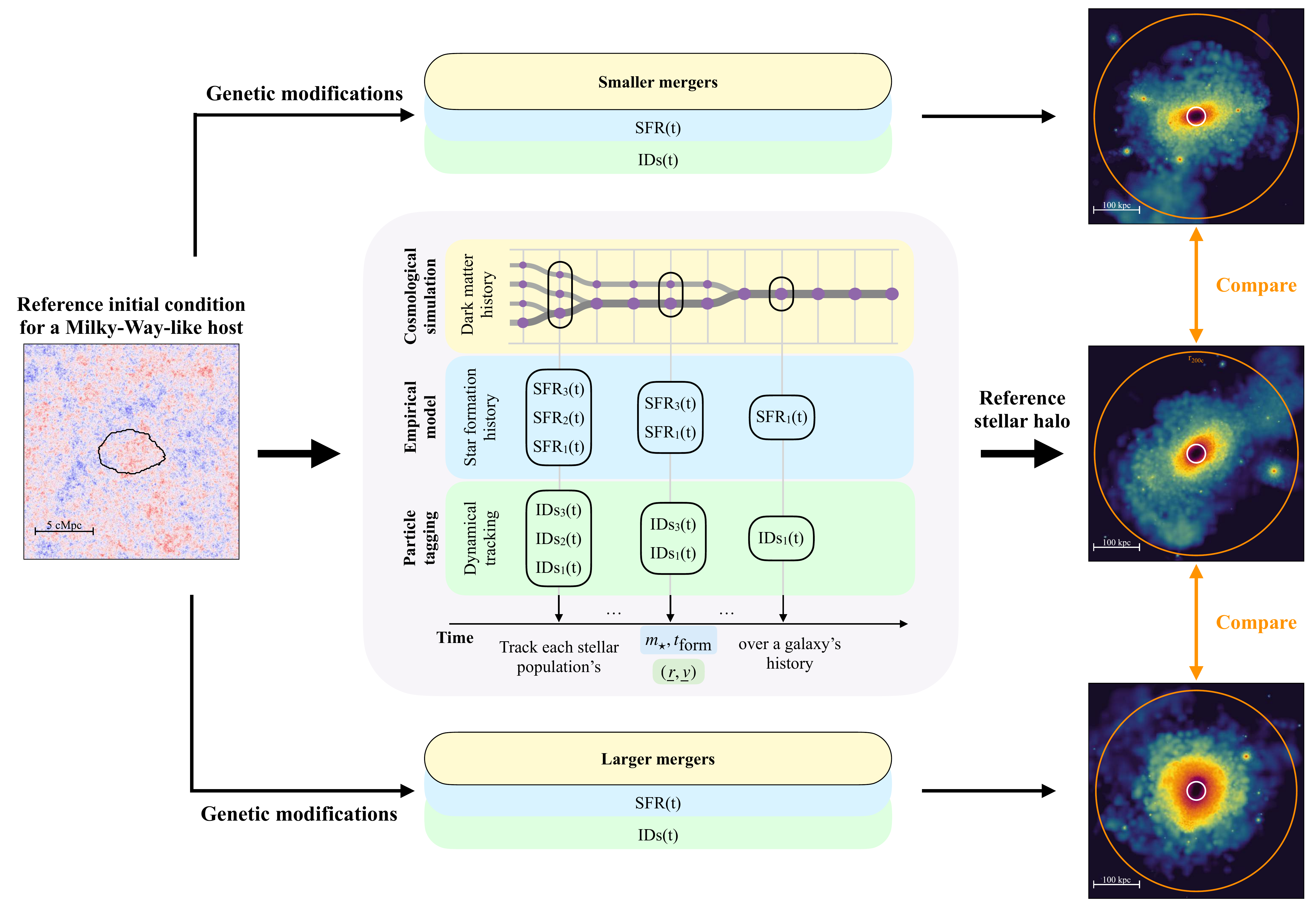}

    \caption{Overview of how we construct stellar haloes for each genetically modified dark matter host. Starting from a cosmological initial condition, we use a zoomed, gravity-only simulation to construct the host's merger tree (yellow, Section~\ref{sec:sec:suite}). At each simulation timestamp, we determine star formation rates for all progenitors over the age of the Universe using the empirical \textsc{emerge} model (blue, Section~\ref{sec:sec:emerge}) and select a set of dark matter particles to represent the phase space of newborn stars following the `particle tagging' framework (green, Section~\ref{sec:sec:cooper}). Tracking each stellar population to $z=0$ allows us to reconstruct the diffuse halo of stars in a galaxy's outskirts (stellar density map, right). We further use genetic modifications to engineer specific, targeted changes to the host's cosmological merger history (top and bottom arrows, and Figure~\ref{fig:halo715_mah} and~\ref{fig:halo685_mah}), enabling a controlled and flexible approach to explore how the interplay between cosmological merger histories (Figure~\ref{fig:mshalo} and~\ref{fig:radialprofiles}) and galaxy formation physics (Figure~\ref{fig:dwarfsfhsvariations} and~\ref{fig:fmbvariations}) shapes galactic outskirts.
    }
    \label{fig:overview}
\end{figure*}

We first introduce our methodology to probe the role of merger histories in the formation of stellar haloes. We describe how we construct stellar haloes for each host's merger history, complementing cosmological dark matter-only simulations (Section~\ref{sec:sec:suite}) with empirical star formation histories (Section~\ref{sec:sec:emerge}) and a `particle tagging' approach (Section~\ref{sec:sec:cooper}). The specific objects presented in this work, together with their genetically modified mass growth histories are characterized in Section~\ref{sec:gms}. Figure~\ref{fig:overview} provides a visual summary of our approach. 

\subsection{Zoomed, dark matter-only cosmological simulations} \label{sec:sec:suite}

We first generate a dark matter-only cosmological volume with a box size $50  \, \hMpc \approx 73 \, \Mpc$ and mass resolution $\mdm = \xMsol{1.2}{8}$, from which we select all objects presented in this work. All simulations are started from cosmological initial conditions evolved to $z=99$ using first-order linear theory (\citealt{Zeldovich1970}) and with parameters $\Omega_{m} = 0.3139$, $h = 0.6727$, $\sigma_8 = 0.8440$, $n_s = 0.9645$ compatible with \citet{PlanckCollaboration2016} (table 3, column 4).

We select isolated haloes, i.e. with no more massive neighbours within $5 \, \rvir$, where $\rvir$ is the radius enclosing 200 times the critical density of the Universe. In our goal to target Milky Way-mass objects, we further pick haloes with $\xMsol{7.5}{11} \leq \Mvir \leq \xMsol{1.5}{12}$ (e.g. \citealt{ Watkins2019, Cautun2020, Deason2021}; see \citealt{Wang2020} for a review), where $\Mvir$ is the mass enclosed within $\rvir$. We track the region enclosed by $3 \, \rvir$ at $z=0$ back to $z=99$ and produce zoomed initial conditions refining the mass resolution within this region to $\mdm = \xMsol{2.3}{5}$ using the \textsc{genetIC} software (\citealt{Stopyra2021}). This region is large enough to alleviate contamination by low-resolution particles, even after modifying mass accretion histories by performing genetic modifications (see Section~\ref{sec:gms}). We verified that in all objects, low-resolution particles contribute at most 0.1 per cent of the total particle number. We further apply the procedure in \citet{Pontzen2021} to all initial conditions to strongly damp the overall streaming of the Lagrangian region within its parent volume and mitigate advection errors during integration.

We then follow the gravitational collapse of dark matter to $z=0$ using the \textsc{ramses} adaptive mesh refinement code (\citealt{Teyssier2002}), solving the Poisson equation through a multi-grid approach (\citealt{Guillet2011}). We allow refinement following a Lagrangian strategy when 8 dark matter particles share a given cell, up to a maximum spatial resolution $\softening = 140 \, \pc$. We identify haloes using the \textsc{hop} halo finder (\citealt{Eisenstein1998}), and construct merger trees matching haloes between timesteps (Figure~\ref{fig:overview}, yellow boxes) using the \textsc{pynbody} and \textsc{tangos} libraries (\citealt{Pontzen2013, Pontzen2018}).

\subsection{Empirical star formation histories}  \label{sec:sec:emerge}

From the constructed dark matter-only merger tree, we next need to assign star formation histories to each progenitor of the host halo (Figure~\ref{fig:overview}, blue boxes). We implement an approach following the empirical model \textsc{emerge} (\citealt{Moster2018, OLeary2021}), that ties a star formation rate to a dark matter halo through
\begin{equation}
  \label{eq:sfr}
   \SFR (\Mvir, t) = \baryonfraction \ \efficiency (\Mvir, t) \ \Mgrowth  \, , 
\end{equation}
where $\Mgrowthinline$ is the growth of a halo's virial mass averaged over the dynamical time $\tdyn = \sqrt{\rvir^3/ (\G \Mvir)}$, $\baryonfraction$ is the universal baryon fraction and $\efficiency$ encodes the instantaneous efficiency at which baryons are turned into stars.

We adopt the same parametrization and functional form for $\efficiency (\Mvir, t)$ as in \citet{Moster2018}, eq.~5:
\begin{equation}
  \label{eq:efficiency}
  \efficiency(\Mvir, t) = 2 \, \effnorm (t) \, \Bigg [ \Big ( \frac{\Mvir}{\Mone (t)} \Big )^{- \lmslope(t)} + \Big (\frac{\Mvir}{\Mone(t)} \Big )^{\hmslope(t)} \Bigg]^{-1} \, ,
\end{equation}
where $\Mone$ is the instantaneous pivot mass separating the high-mass regime with slope $\hmslope$ and low-mass regime with slope $\lmslope$, and $\effnorm$ normalizes the expression. We further adopt the same parametrization with redshift for $\effnorm$, $\Mone$, $\lmslope$ and $\hmslope$ from \citet{Moster2018}, eq.~7~--~10:
\begin{align}
  \label{eq:efficiency_redshift}
  \effnorm (z) &=  \effzero +  \effz \, \big( \frac{z}{1 + z} \big)\, ,\\
  \label{eq:pivot_redshift}
  \Mone (z) &=  10^{\Mzero +  \Mz \, (\frac{z}{1 + z})} \, \Msol ,\\
  \label{eq:lmslope_redshift}
  \lmslope (z) &=  \lmslopezero +  \lmslopez \, \big( \frac{z}{1 + z} \big) \, ,\\
  \label{eq:hmslope_redshift}
  \hmslope (z) &=  \hmslopezero +  \hmslopez \, \big( \frac{z}{1 + z} \big) \, .
\end{align}

Star formation histories can then be integrated over time to build up a galaxy's stellar mass. We sum the stellar mass formed over time in all branches of the merger tree, accounting for the mass loss of stellar populations as a function of age according to \citet{Moster2018}, eq. 12. In addition to the \textit{in-situ} component within the main progenitor, we track all merger events recorded in the merger tree and aggregate their stellar masses into an accreted component. The sum of both components defines the total stellar mass of the galaxy, $\Mstar$.

In typical applications of empirical models (e.g. \citealt{Moster2018,Behroozi2019}), the parametrization defined through Equations~\ref{eq:sfr} to \ref{eq:hmslope_redshift} is applied to a statistical population of dark matter haloes and the maximum-likelihood parameters of the model (e.g. $\Mzero$ and $\Mz$) are then determined by fitting against observed properties of the galaxy population. However, our aim is to produce star formation histories for a handful of genetically modified, high-resolution haloes, rather than a population of lower-resolution objects. We therefore do not attempt to re-fit $\Mzero$, $\Mz$, $\effzero$, $\effz$, $\lmslopezero$, $\lmslopez$, $\hmslopezero$ and $\hmslopez$, but rather compute the virial mass growth of every progenitor at each simulation timestep and convert it into a star formation rate with the best-fitting parameters determined by \citealt{OLeary2021}, table 1 (reproduced in Table~\ref{table:bestfitparams}). We show in Appendix~\ref{app:smhm} that our obtained relationship between stellar mass and halo mass is consistent with \citet{OLeary2021}, \MR{confirming the consistency of extrapolating} empirical best-fitting parameters determined from a statistical population to a handful of Milky Way-mass zoom simulations (see also \citealt{Wang2021}).

Furthermore, since we aim to explicitly track the dynamics of stellar populations during mergers to build stellar haloes using a `particle tagging' approach (Section~\ref{sec:sec:cooper}), a prescription describing the evolution of satellite haloes once they infall their host (e.g. section 3.2 in \citealt{Moster2018}) would potentially obscure the pure signature of the accretion history. We thus assume that \textit{in-situ} star formation ceases at a satellite's first infall and tracks the subsequent dynamical evolution explicitly within the `particle tagging' framework. Although star formation in satellites can proceed after their virial crossing (e.g. \citealt{Tollerud2011, Wetzel2012, Akins2021}) or be re-ignited by interactions (e.g. \citealt{Wright2019, DiCintio2021}), strong uncertainties remain in treating low-mass satellite quenching within \MR{semi-empirical models (e.g. \citealt{Kravtsov2021, Wang2021})}. Further investigations are hence required to include stellar ages and metallicity distributions in our analysis, but we leave them to future work since we focus here on the distribution of mass in the stellar halo and its relationship with galaxy merger histories.

\subsection{Particle tagging dynamical histories} \label{sec:sec:cooper}

We next post-process our simulations with a `particle tagging' approach to populate the stellar halo within each dark matter host. Particle tagging selects a carefully chosen set of dark matter particles extracted from the gravity-only simulation to represent the phase-space evolution of stars over time (e.g. \citealt{Bullock2005, DeLucia2008, Cooper2010, Tumlinson2010, Libeskind2011, Rashkov2012, Amorisco2017Atlas}). Here, we re-implement the \textsc{stings} particle tagging framework (\citealt{Cooper2010, Cooper2017}) which we summarize next. \textsc{stings} flags dark matter particles for tagging regularly over the age of the Universe, at every simulation snapshot (here every $\approx 100\, \Myr$), which is key to accurately follow the phase-space diffusion of newborn stars and reproduce stellar haloes from hydrodynamical simulations (\citealt{LeBret2017}).

At each merger tree timestep, we order dark matter particles within every progenitor by their binding energy and flag a fixed fraction of the most bound dark matter particles, $\fmb$, to represent the dynamics of newly formed stars (Figure~\ref{fig:overview}, green boxes). To re-construct the stellar halo at a given time, we walk through the merger tree, identifying the position and velocities of dark matter particles within the instantaneous set of `tagged' particles in each progenitor. We then construct a stellar population for each particle within the set, with phase space matching its dark matter counterpart and a stellar mass $m_{\star}$ derived from the instantaneous star formation rate $\text{SFR}(t)$ through
\begin{equation}
  m_{\star} (t) = \frac{\text{SFR}(t) \, \Delta t \, \mdm}{ \fmb \, \Mvir(t)} \, ,
  \label{eq:mstarperdmparticle}
\end{equation}
where $\Delta t$ is the time interval between two simulation snapshots. With this procedure, the same dark matter particle can be tagged on multiple occasions over the merger tree, accumulating its own star formation history. To avoid concentrating stellar populations at the same spatial positions, we displace them by a random perturbation uniformly drawn from $[-\softening, \softening ]$, the simulation's maximum resolution. We verified that different random draws from this distribution are indistinguishable from results presented in Section~\ref{sec:stellarhaloes}, as are draws from a distribution spawning an interval of twice or three times the simulation's spatial resolution. 

The sole free parameter in this approach is $\fmb$, encoding how tightly bound newly formed stars are within a halo. This `bound fraction' thus effectively defines the extent of the star-forming region and strongly relates to galaxy sizes when integrated over time (\citealt{Cooper2013}), which can in turn be used to calibrate it. The inferred best-fitting value, however, depends on both the star formation history model and the observables used for calibration --- from $\fmb = 1\%$ (\citealt{Cooper2010, Rashkov2012}) when comparing to dwarf galaxies structural properties, to $1 \% \leq \fmb \leq 3\% $ for more massive galaxies (\citealt{Cooper2013}), to $\fmb = 5\%$ when matching to a hydrodynamic simulation (\citealt{LeBret2017, Cooper2017}). Given this range of allowed values, we adopt $\fmb = 2 \%$ as our fiducial value and systematically explore how varying it affects the stellar halo in Section~\ref{sec:boundfraction}.

\section{Genetically modified Milky Way-mass hosts} \label{sec:gms}

We next describe how we construct two families of `genetically modified' initial conditions for Milky Way-mass dark matter haloes. Starting from a reference cosmological initial condition, genetic modifications create different versions of the same host halo, each with carefully crafted modifications to its past merger history (Figure~\ref{fig:overview}, top and bottom arrows). This enables controlled studies by comparing how each merger scenario affects a galaxy's final observables and structural properties, in our case the stellar halo (Figure~\ref{fig:overview}, right images, see Section~\ref{sec:stellarhaloes} for results).  

\begin{table}
  \centering
  \caption{Genetic modifications to obtain our two families of modified mass growth 
  histories (Figure~\ref{fig:halo715_mah} and Figure~\ref{fig:halo685_mah}). We adjust the mean density of the Lagrangian region (second column) to ensure matching dynamical masses at $z=0$. To modify merger histories (third column), we target the local variance within the Lagrangian region for the \lateformer family, and the linear overdensity enclosing the last major merger for the \mwlike family. All changes are quoted relative to the reference initial condition in each family.}
  \label{table:runs}

  \setlength\tabcolsep{4.0pt}
     \begin{tabular}{c c c c }
     \hline
     Simulation name & GM: Dynamical & GM: Merger  \\
     & Mass at $z=0$ (\%) & History (\%) \\
     \hline
     \hline
    Reference: Late former & $=$ & $=$ \\
    GM: Earliest major mergers & $=$ & $ + 20$ \\
    GM: Earlier major mergers & $=$ & $ + 10$ \\
    GM: Later major mergers & $=$ & $ - 10$ \\
    GM: Latest major mergers & $=$ & $ - 20$ \\
    \hline
    \hline
    Reference: Milky Way-like & $=$ & $=$ \\
    GM: Smallest last major merger  & $ = $ & $ - 10$ \\
    GM: Smaller last major merger & $ = $ & $ - 5$ \\
    GM: Larger last major merger & $ + 2.5$ & $ + 10$ \\
    GM: Largest last major merger & $ + 5$ & $ + 20$ \\
    \hline
    \end{tabular}
\end{table}

\subsection{Altering cosmological merger histories} \label{sec:sec:families} 

\begin{figure}
  \centering
  \includegraphics[width=\columnwidth]{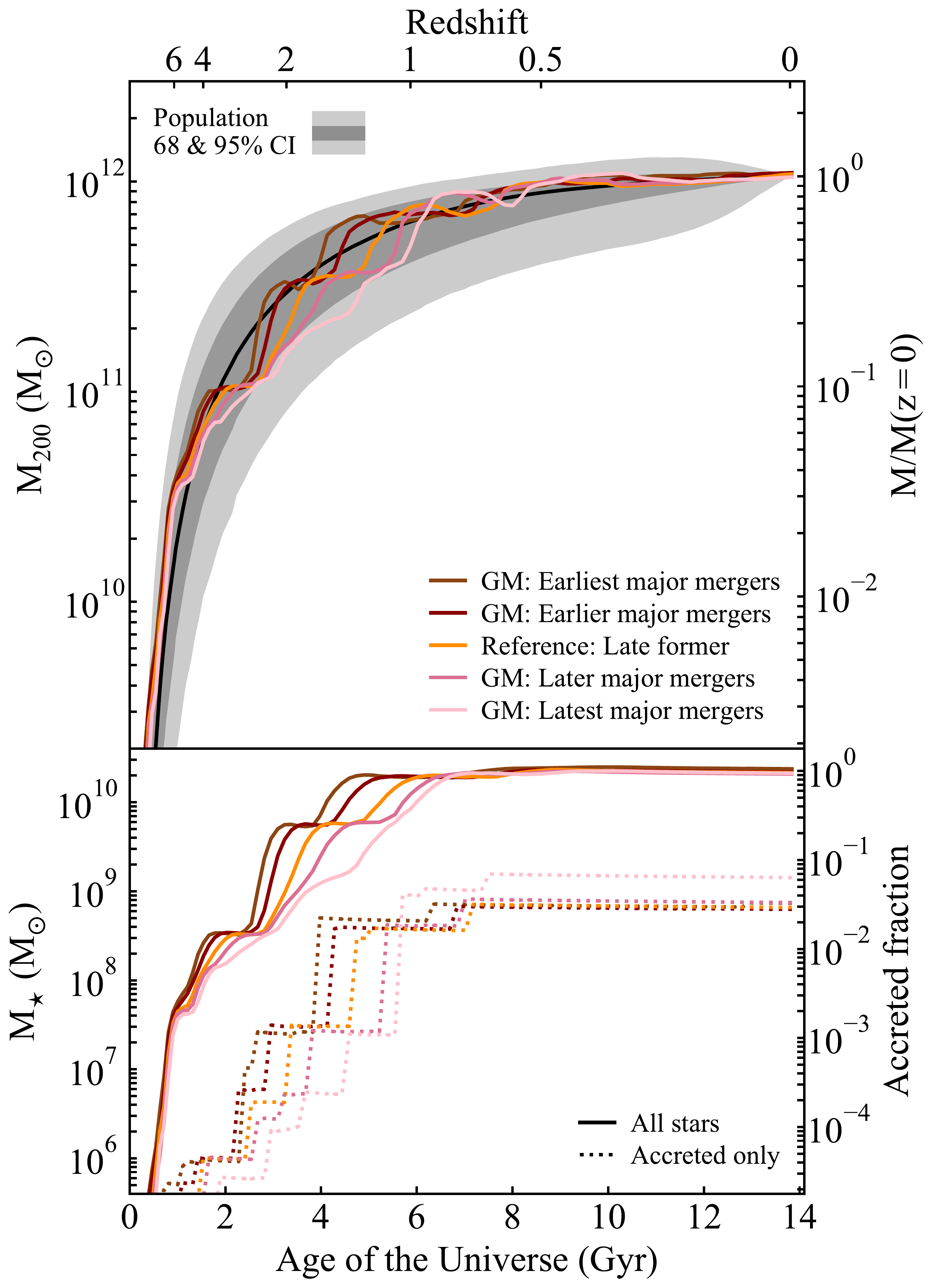}
  \caption{Growth of halo mass (top) and stellar mass (bottom) \MR{with cosmic time} for the \lateformer family of initial conditions. We target the late mergers in the reference object (orange) and construct genetically modified initial conditions to make their mass delivery increasingly earlier (red and brown) and later (magenta, pink). The resulting family of histories samples a range of late forming objects compared to the overall halo population (grey contours). The build-up of the total stellar mass (bottom, solid curves) smoothly tracks the dark matter mass growth history, as expected for empirical galaxy formation modelling.  Different merger scenarios, however, modify the timing and amplitude of the mass delivery in the accreted component (dashed), allowing us to compare the respective response of stellar halo observables (Figure~\ref{fig:mshalo} and~\ref{fig:radialprofiles}).
  }
  \label{fig:halo715_mah}
\end{figure}

\MR{Using the procedure described in Section~\ref{sec:sec:suite}, we first pick nine isolated Milky Way-mass host dark matter haloes, for which we construct zoomed initial conditions (presented in Appendix~\ref{app:randomsuite}). We then re-select two reference objects as starting points for our genetically modified suites, with contrasting growths over the age of the Universe while remaining typical objects within our selection suite. 

We choose one object that assembles late relative to the overall population, with significant mergers at $z\leq2$ (\textit{Late Former}; Figure~\ref{fig:halo715_mah}), and one that assembles early, with a last major merger at $z\approx2$. This choice connects to future goals to interpret galactic observations, as this merger history approximately resembles our Milky Way's history (\textit{Milky Way-like}; Figure~\ref{fig:halo685_mah}).} We assess the abundance of mass accretion histories by comparing our objects to a halo population extracted from the IllustrisTNG simulation suite (\citealt{Naiman2018, Springel2018, Nelson2018, Marinacci2018, Pillepich2018, Nelson2019}), specifically from `TNG-300-1-Dark' with a volume of $(300 \, \Mpc)^3$ and $\mdm = \xMsol{7.7}{7}$. We select the 28,475 central dark matter haloes with final virial masses $\xScientific{8}{11} \leq \Mvir \leq \xMsol{2}{12} $, extract their fractional mass growth over time using the provided \textsc{sublink} merger trees (\citealt{Rodriguez-Gomez2015})\footnote{We verified that using alternative \textsc{lhalotree} merger trees (\citealt{Springel2005}) yield comparable results.} and compute median and confidence intervals at each cosmic time rescaled to match the final mass of the halo of interest (grey contours in Figure~\ref{fig:halo715_mah} and Figure~\ref{fig:halo685_mah}). 

From our two reference objects, we then construct families of alternative, genetically modified initial conditions altering their merger histories (\citealt{Roth2016, Rey2018, Stopyra2021}). Table~\ref{table:runs} lists the modifications made in this work which we describe in details now. 

\textit{Late-former family} -- We start from a reference halo with a final dynamical mass of $\Mvir = \xMsol{1.08}{12}$, which experiences a series of three significant mergers happening late in its assembly, with first infalls between $z=2$ and $z=0.7$. Being late-forming, all histories in this family are expected to yield an extended, massive stellar halo (see e.g. \citealt{Elias2018, Monachesi2019}). We target this series of mergers using `variance' genetic modifications (\citealt{Rey2018}), following the procedure in \citet{Rey2019VarianceDMOs}. Merger progenitors have masses $\approx \xMsol{9}{10}$ and we thus modify the variance in the Lagrangian region on scales of $0.50 \, \hMpc$. We smooth (quicken) the mass delivery from these mergers by gradually reducing (increasing) the variance by up to 20 per cent while maintaining the mean density of the Lagrangian region to ensure matching halo masses at $z=0$. The resulting family of mass growths is shown in Figure~\ref{fig:halo715_mah}, and spans a range of late-forming histories compared to the overall population. 

\textit{Milky Way-like family} -- We additionally select an object designed to resemble the Milky Way's mass accretion history, with a final dynamical mass of $\Mvir = \xMsol{9.98}{11}$, a merger history dominated by a major event with first infall around $z\sim2$ and dark matter mass ratio of 1:3 \MR{compatible with GAIA-Enceladus-Sausage} (e.g. \citealt{Belokurov2018, Helmi2018, Naidu2021GSECharacterisation}), and a quiet history afterwards with no mergers with ratios greater than 1:11. We target this last major merger, tracking the particles composing it at $z=2$ to the initial conditions and increasing (decreasing) the density in this sub-region of the Lagrangian patch to make the merger more (less) significant (\citealt{Roth2016, Pontzen2017}). The global mean density in the patch is adjusted to ensure matching final masses at $z=0$. The resulting family of mass accretion histories exists of a range of early-forming histories compared to the population median, and is shown in Figure~\ref{fig:halo685_mah}.

\subsection{Merger history alterations and galaxy stellar masses} \label{sec:sec:mergers}

\begin{figure}
  \centering
    \includegraphics[width=\columnwidth]{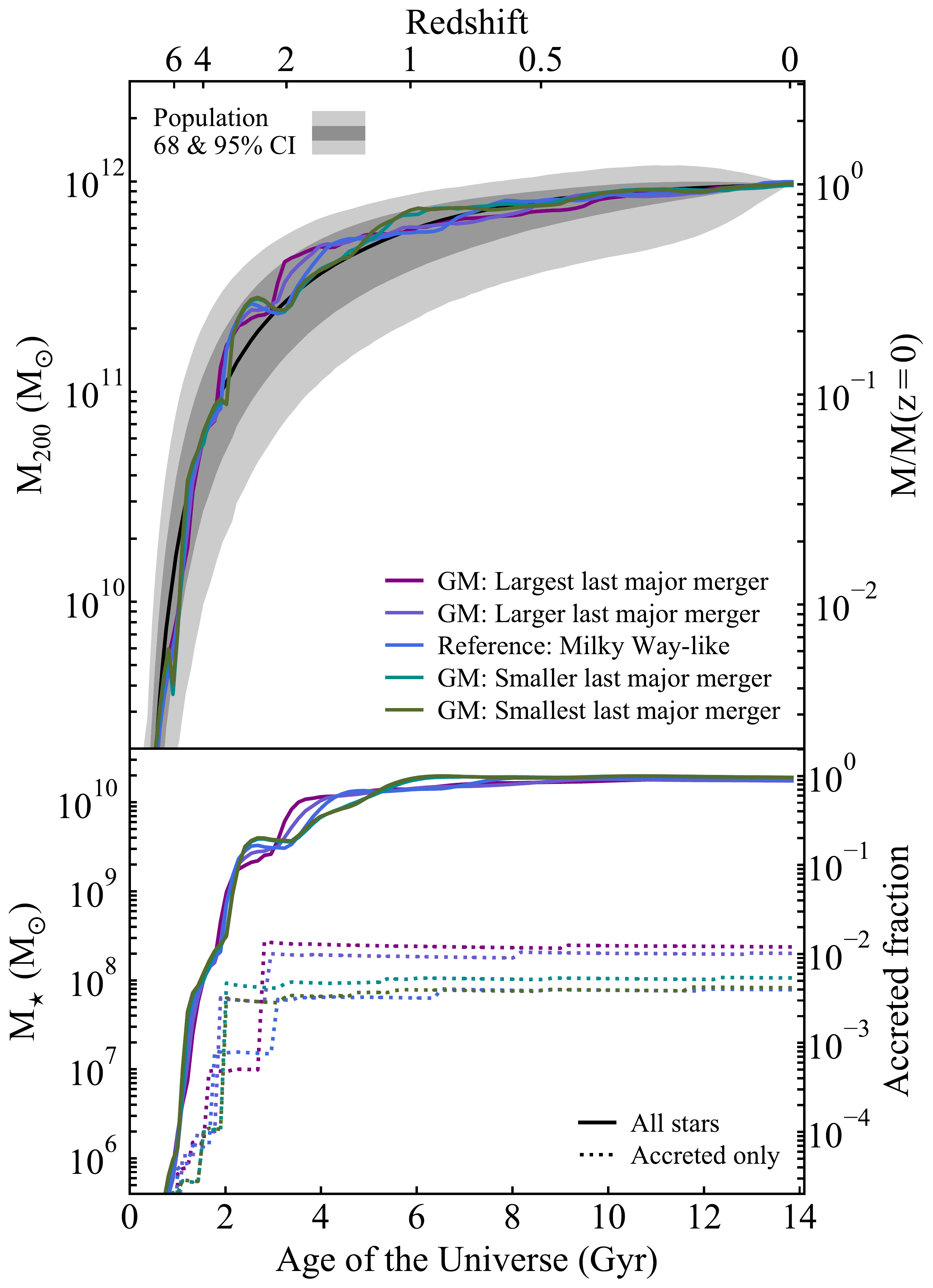}

    \caption{Same as Figure~\ref{fig:halo715_mah}, but modifying the last major merger at $z\sim2$ of a dark matter halo with an early forming, Milky Way-like, merger history. We single out this event, making it, in turn, increasingly larger (purple, navy) and smaller (cyan, green) compared to the reference case (blue). The family spans a range of early-forming histories compared to the overall population (grey contours).
    }
    \label{fig:halo685_mah}
\end{figure}

Figure~\ref{fig:halo715_mah} and Figure~\ref{fig:halo685_mah} show the response of the central galaxy's stellar mass build up (bottom) to our alterations to the halo mass growth history (top) for the \lateformer and \mwlike families respectively. In all objects, the total stellar mass growth (solid) smoothly tracks its corresponding dynamical mass history, as expected from tying \textit{in-situ} star formation to halo mass growth in an empirical galaxy formation model (Section~\ref{sec:sec:emerge}). All galaxies have present-day total stellar masses within the one-sigma scatter predicted by \textsc{emerge}, $\Mstar (\Mvir = 10^{12} \, \Msol) = 2.8^{+1.3}_{-1.0} \, 10^{10} \, \Msol$, validating the self-consistency of our star formation histories (see also Appendix~\ref{app:smhm}). 

The tight grouping observed in final $\Mstar$ -- all modified objects are within ten per cent of their family’s reference galaxy -- is a key feature of this study. It allows us to isolate the role of merger histories in building  the stellar halo at fixed final dynamical \textit{and} total stellar mass budget. This grouping arises from the coupling between the empirical galaxy formation model and the minimal nature of the `genetic modification' algorithm (\citealt{Roth2016, Rey2018}). In empirical models the scatter in stellar mass at fixed halo mass is generated through diversity in the halo formation histories and environments \citep{Moster2018, Behroozi2019}, and similar histories at fixed halo mass thus result in similar stellar masses. While the genetic modifications technique modifies targeted aspects of the merger history, each new scenario is ensured to resemble its reference as closely as possible in all other, untargeted features. Here, all galaxies hence have similar large-scale structures (e.g. \citealt{Rey2019VarianceDMOs}), early ($z\geq4$) and late ($z\leq0.5$) formation, resulting in minimal deviations in $\Mstar$ at $z=0$ for all of our modified histories (Figure~\ref{fig:halo715_mah} and Figure~\ref{fig:halo685_mah}). In this work, we capitalize on these minimal deviations, to fix an important control variable, $\Mstar$, in the build-up of the stellar halo, but we stress that our modelling may not capture the full diversity of galaxy stellar masses. The simplifying assumption inherent to empirical models, that the averaged dynamical mass growth drives the response of the central's star formation history, neglects rapid, merger-induced internal events such as starbursts (e.g. \citealt{Renaud2014, Sparre2016}) or star formation quenching (e.g. \citealt{Pontzen2017}). Such mechanisms may introduce diversity in the final $\Mstar$ as well as other galaxy and halo properties. We will re-simulate our initial conditions with hydrodynamical simulations to explore how these merger-induced mechanisms correlate with the build-up of stellar haloes in future work.

In contrast to the smooth, total stellar mass growth, the accreted component (bottom, dashed lines) shows discrete steps corresponding to individual accretion events which are modified in amplitude and timing by our genetic modifications. Again, several galaxies converge to similar accreted fractions (bottom right axis) due to compensation effects inherent to the minimal nature of genetic modifications: because we enforce the same final dynamical mass, growing given structures is compensated by diminishing others. For moderate adjustments in the progenitors' dynamical masses and times of accretion, this leaves the total accreted sum unaffected. An example can be observed in the \lateformer family (Figure~\ref{fig:halo715_mah}, bottom), for which all family members but one converge to nearly identical accreted fractions. Despite these identical accreted fractions, we will show that differences in the timing and prominence of individual events lead to markedly different stellar haloes (Section~\ref{sec:stellarhaloes}).

The notable exceptions to these compensation effects are the `GM: Latest major mergers' (pink, Figure~\ref{fig:halo715_mah}), `GM: Larger last major merger' and `GM: Largest last major merger' (navy and purple in Figure~\ref{fig:halo685_mah} respectively) galaxies. We verified that these larger, uncompensated changes occur when genetic modifications cross a `tipping point', altering the topology of the reference halo's merger tree (i.e. the number and order of mergers) compared to smoothly varying merger ratios and timings (see \citealt{Rey2019VarianceDMOs} for further discussion). For instance, in the `Reference: Late former' galaxy, two of our targeted late mergers are incorporated one by one in the main progenitor, which experiences two events. By contrast, in the `GM: Latest major mergers' galaxy, these two objects combine first together, before merging with the main progenitor which experiences only one event. Due to this recombination, the last merger remnant is larger at infall (1.6 times higher halo mass and 2.1-times higher stellar mass than the biggest event in the reference run) and drives the increase in accreted fraction. We will see in Section~\ref{sec:stellarhaloes} that such merger tree reconfigurations lead to `jumps' from otherwise smoothly evolving trends in stellar halo properties.

Finally, before we turn to studying the response of each galaxy's stellar halo to our modifications, there are potential sources of purely numerical variations that we wish to exclude. The first is numerical resolution which, even for gravity-only simulations, can be insufficient to accurately capture the dynamics and stripping of small progenitors contributing to the stellar halo (e.g. \citealt{vandenBosch2018}). Secondly, numerical simulations of structure formation are intrinsically chaotic, introducing run-to-run variance in the properties of each object (\citealt{Genel2019, Keller2019}) which could obscure trends obtained from genetic modifications. We show in Appendix~\ref{app:numerics} that variance from both effects is subdominant compared to variations arising from our alternative merger histories.

\section{The imprint of cosmological merger histories on stellar haloes}  \label{sec:stellarhaloes}

\subsection{Generating diversity in stellar halo masses} \label{sec:sec:smasses}

\begin{figure}
  \centering
    \includegraphics[width=\columnwidth]{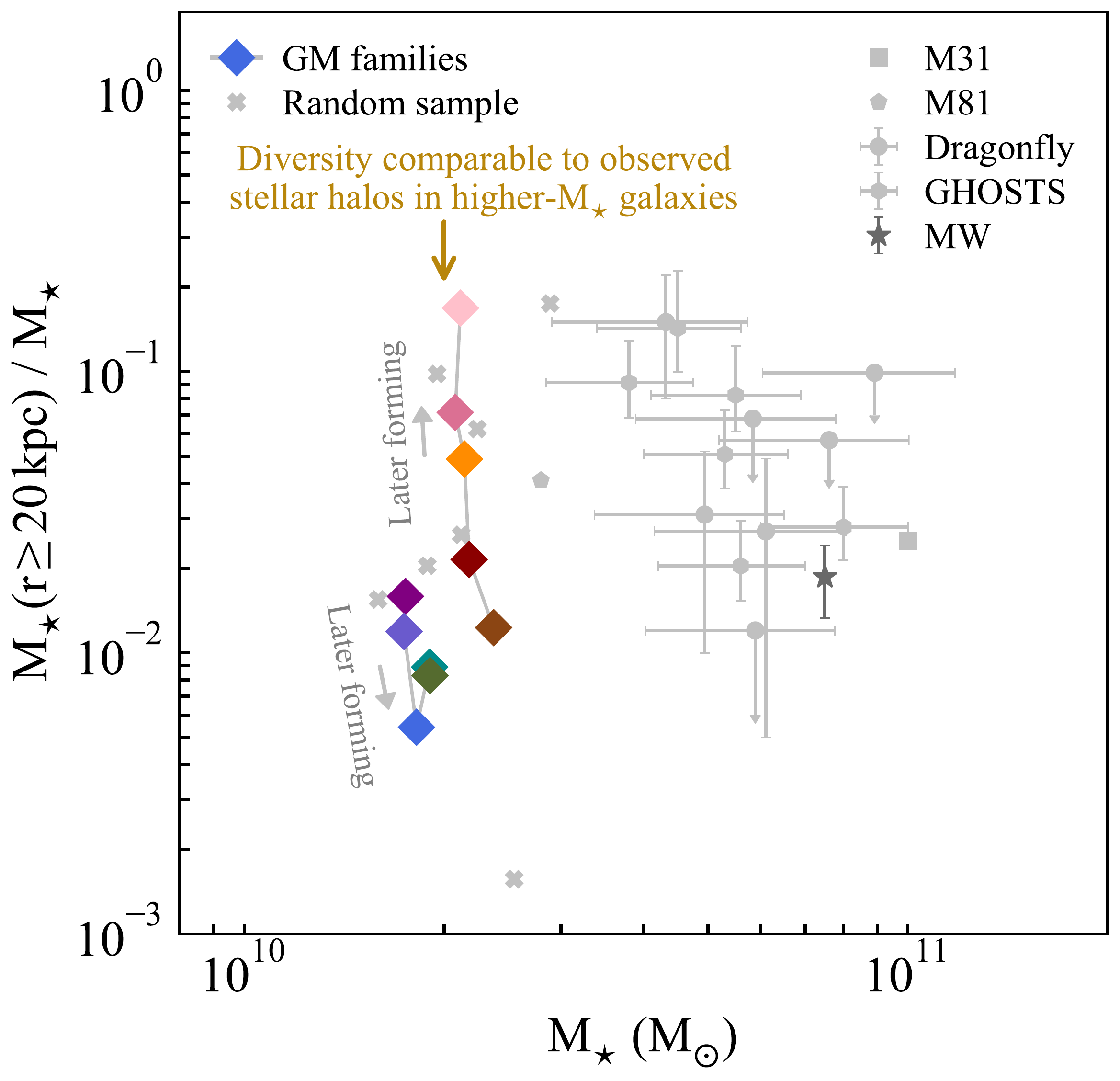}

    \caption{\MR{Fraction of stellar mass in the halo (defined as the fraction of mass outside 20 kpc) against total stellar mass as we vary cosmological merger histories in our two families of genetically modified, initial conditions (individual colours matching Figure~\ref{fig:halo715_mah} and~\ref{fig:halo685_mah})}. Both families have similar dynamical masses and total stellar masses at $z=0$, but order-of-magnitude variations in stellar halo fraction (vertical spread). Systematically varying merger scenarios for two dark matter haloes generate comparable diversity to the observed stellar haloes in Local-Volume galaxies (grey points): \MR{M31 \citep{Courteau2011}, the \textsc{dragonfly} \citep{Merritt2016} and \textsc{ghosts} samples \citep{Monachesi2016}, the Milky Way \citep{Deason2019} and M81 \citep{Smercina2020}. Genetically-modified merger scenarios also span a comparable range in stellar halo mass diversity when compared to a suite of randomly-selected simulated objects (crosses; further discussed in Appendix~\ref{app:randomsuite}).}
    }
    \label{fig:mshalo}
\end{figure}

We now turn to describing how modifications to merger histories affects the properties of stellar haloes around Milky Way-like galaxies. We focus first on bulk, integrated properties such as the total stellar mass in the halo. From the post-processed, tagged simulation snapshot, we centre on the stellar component using the shrinking sphere algorithm (\citealt{Power2003}) and compute the three-dimensional spherical density profile in 150 bins evenly spaced in log radius between 1 and 300 kpc. (We verified that switching to 100 and 200 bins, or to linearly spaced bins leaves our conclusions unchanged). We do not attempt to remove satellite galaxies from individual profiles at $z=0$, causing the profiles to be noisier in the host's outskirts where bound objects can dominate the light profile, but ensuring a complete account of the accreted stellar mass. Integrating the density with radius, we show in Figure~\ref{fig:mshalo} the fraction of stellar mass in the stellar halo defined as the enclosed stellar mass outwards of $20\, \kpc$ divided by the total stellar mass. We verified that varying our spatial definition to 10 and 30 kpc modifies the absolute values of stellar halo mass ratios, but not the relative trends observed between family members (nor any results in Sections~\ref{sec:dwarfs} and~\ref{sec:boundfraction}).

To benchmark variations due to genetic modifications, we qualitatively\footnote{We verified that projecting our stellar haloes along a random line-of-sight conserves relative trends between family members but do not attempt to correct for varying stellar halo definitions and observing techniques between studies (e.g. \citealt{Sanderson2018}).} compare our two families with observational stellar halo fractions in Local Volume, Milky-Way analogues (grey symbols in Figure~\ref{fig:mshalo}): M31 \citep{Courteau2011}, the \textsc{dragonfly} \citep{Merritt2016} and \textsc{ghosts} samples \citep{Monachesi2016}, the Milky Way \citep{Deason2019} and M81 \citep{Smercina2020}. \MR{We further show stellar haloes from seven, randomly-selected Milky Way-mass hosts (crosses) from which the original reference cases of our genetically-modified families are selected (Appendix~\ref{app:randomsuite}). Each stellar halo in this sample is constructed using the same pipeline as for genetic modifications, providing a controlled comparison albeit with small-number statistics. We contrast this in Appendix~\ref{app:tng} using a large sample of galaxies obtained from `TNG300-1', but with differing galaxy formation modelling.}

Varying merger histories introduces a strong diversity in stellar halo mass fraction, with a scatter ranging from 0.44 to 1.1 dex across the \mwlike and \lateformer families respectively, combining to 1.5 dex across all objects. Within a family, all galaxies have nearly identical large-scale structure, final dynamical and total stellar mass due to the minimal nature of genetic modifications (Section~\ref{sec:gms}). The vertical spread in Figure~\ref{fig:mshalo} therefore cleanly probes the expected scatter in stellar halo mass ratios generated by the diversity of possible cosmological histories for a galaxy, spreading across the full extent in observed stellar haloes \MR{and randomly-selected simulated hosts}. The systematically higher stellar masses of observed central galaxies result from the observational selection of Milky Way analogues with disc morphologies and $\Mstar \approx \xMsol{7}{10}$, which are significantly overmassive compared to the median $\Mstar \approx \xMsol{3}{10}$ predicted by empirical models at $\Mvir \approx \xMsol{1}{12}$ for our simulated objects (see e.g. \citealt{Posti2019, Posti2021} for further discussion).

In addition to this exposed diversity, our experiments allow us to test the physical nature and scatter in correlations between the assemblies of dark matter and stellar haloes, e.g. that later assembling dark matter haloes are correlated with more massive stellar haloes across the galaxy population (\citealt{Elias2018}). Dark matter haloes in the \lateformer family are systematically later forming, achieving half of their final $\Mvir$ at $\zform = 1.42 \rightarrow 1.40 \rightarrow 1.18 \rightarrow 1.07 \rightarrow 0.99$ from the `GM: Earliest major mergers' (brown) to the `GM: Latest major mergers' galaxies (pink). This ordering leads to growing stellar halo masses (grey arrow in Figure~\ref{fig:mshalo}), tracking the established correlation in the overall population. 

However, our results also highlight the challenges in utilizing stellar halo fractions to infer a galaxy's past merger history, even for one-point summary statistics such as $\zform$. The \mwlike family is systematically earlier forming with genetic modifications, with $\zform = 1.69 \rightarrow 1.64 \rightarrow 1.53 \rightarrow 1.33 \rightarrow 1.31$ from the `GM: Largest last major merger' (purple) to `GM: Smallest last major merger' (green) galaxies. However, diminishing the last major merger in the \mwlike family actually grows an earlier merger ($z\sim4$, which goes from a dark matter mass ratio of 1:2 to 1:1.4 to 1:1.1). Through this compensation, the final `GM: Smaller last major merger' (turquoise) and `GM: Smallest last major merger' (green) acquire an overall higher stellar halo fraction than the reference (visible as the kink in the \mwlike family along the grey arrow in Figure~\ref{fig:mshalo}).

\begin{figure*}
  \centering
    \includegraphics[width=\textwidth]{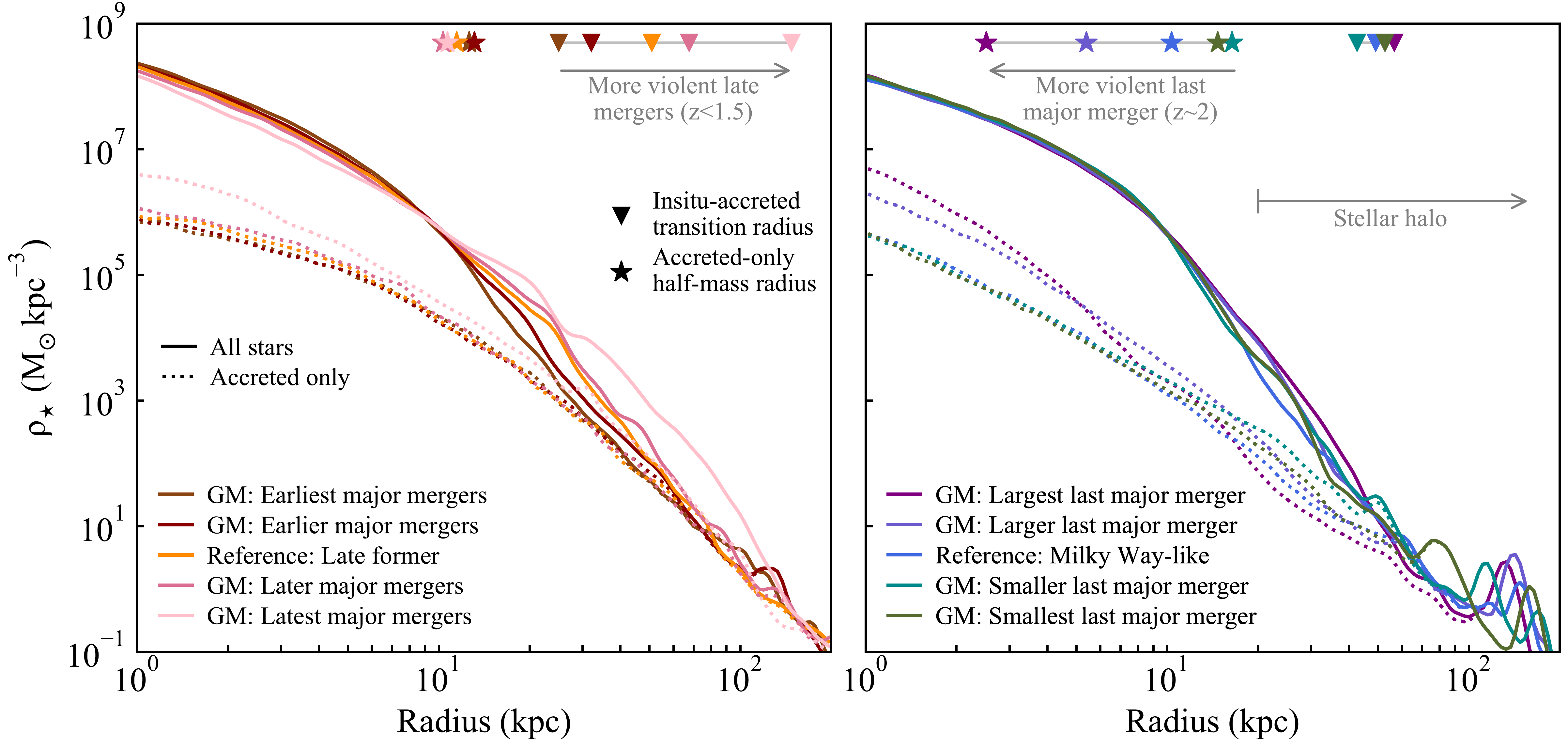}

    \caption{\MR{Spherically-averaged, stellar density profile for the total (solid) and accreted (dotted) components as we vary merger scenarios using the genetic modification technique in the \lateformer (left) and \mwlike (right) families}. Later, more violent mergers in the \lateformer family (left-hand panel, \MR{brown to pink}) increasingly scatter the \textit{in-situ} component into the outskirts, pushing the \MR{radius of} transition from \textit{in-situ} to \textit{ex-situ} dominance outwards (top triangles \MR{and grey arrow}). This drives a strong systematic increase in stellar density in the halo (solid), and thus order-of-magnitude diversity in stellar halo fractions (Figure~\ref{fig:mshalo}). A more prominent last major merger in the early forming \mwlike family (right-hand \MR{panel, green to purple}) sinks deeper due to dynamical friction, making the accreted radial profile (dotted) increasingly steeper and more concentrated (top stars, \MR{grey arrow). We indicate the region defined as the stellar halo (right-hand panel, grey arrow) used to define total masses in Figure~\ref{fig:mshalo}.}
    }
    \label{fig:radialprofiles}
\end{figure*}

In fact, the original, reference Milky Way-like system (blue) is the sole object where both merger events are middle sized -- larger stellar haloes are thus achieved for both earlier and later forming galaxies depending on which of these two mergers is grown. This highlights the sensitivity of the final stellar halo fractions to the entire mass spectrum of progenitors across a galaxy's history, and the challenge in summarizing this information in a one-point measure such as formation time. Studies aiming to establish correlations between dark matter accretion histories and stellar halo assemblies would thus benefit from more detailed metrics than single-number summary statistics (see also \citealt{Wang2020ConcentrationMergers}). We leave the design of such metrics to future work, and now use our ability to compare genetically modified merger scenarios to pinpoint the mechanisms driving diversity in stellar halo structures.

\subsection{Radial structure of stellar haloes} \label{sec:sec:radialstructure}

Figure~\ref{fig:radialprofiles} shows the spherically-averaged, three-dimensional stellar density radial profile for our two families of merger histories. Focussing first again on the \lateformer family (left-hand panel), the observable, total profile (solid) undergoes dramatic structural changes in the stellar halo radial range ($r\geq 20\, \kpc$, grey arrow): for more delayed major mergers, the stellar density systematically increases in the outskirts, matching the increase in stellar halo mass fractions in Figure~\ref{fig:mshalo}. As we delay major merger events, they also become more significant in both infall mass and merger ratios as they have more time to grow outside their host's influence (\citealt{Rey2019VarianceDMOs}). This increasingly violent merger history yields a stronger response of the existing \textit{in-situ} stars, increasing their density in the outskirts. In contrast, the radial stellar density profiles for the accreted stars (dotted lines) are minimally modified.

To quantify the importance of the \textit{in-situ} component in these stellar haloes, we compute the radius at which the \textit{in-situ} radial profile first equals its \textit{ex-situ} counterpart (top triangles). The transition from \textit{in-situ} to \textit{ex-situ} domination is pushed outwards with more violent merger histories (brown to pink, grey arrow), reaching ${\sim} 150 \, \kpc$ for the `GM: Latest major mergers' which is also our most massive stellar halo. In fact, this specific history corresponds to an example `tipping point' in which our genetic modifications have modified the merger tree topology in addition to the merger ratios and infall times (see Section~\ref{sec:gms} for further discussion). This merger tree reconfiguration drives an overall increased accreted budget compared to other family members (Figure~\ref{fig:halo715_mah} and Section~\ref{sec:gms}) but, even at large radii, the radial stellar density profile remains dominated by an \textit{in-situ} component stirred up by this most violent of our merger scenarios.

The re-distribution of existing \textit{in-situ} material outwards is key to explain the order-of-magnitude variations in stellar halo fractions within the \lateformer family (Figure~\ref{fig:mshalo}), with changes to the accreted budget playing a limited role. The dynamical response of existing \textit{in-situ} stars to mergers is thus a crucial driver to the diversity of stellar halo masses and structures, with stronger, more violent, late merging activity making each galaxy's stellar halo more massive (Figure~\ref{fig:mshalo}) but also more \textit{in-situ} dominated (Figure~\ref{fig:radialprofiles}, \MR{see also Appendix~\ref{app:randomsuite} for a discussion of this same mechanism on an independent, randomly-sampled halo)}.

\begin{figure*}
  \centering
    \includegraphics[width=\textwidth]{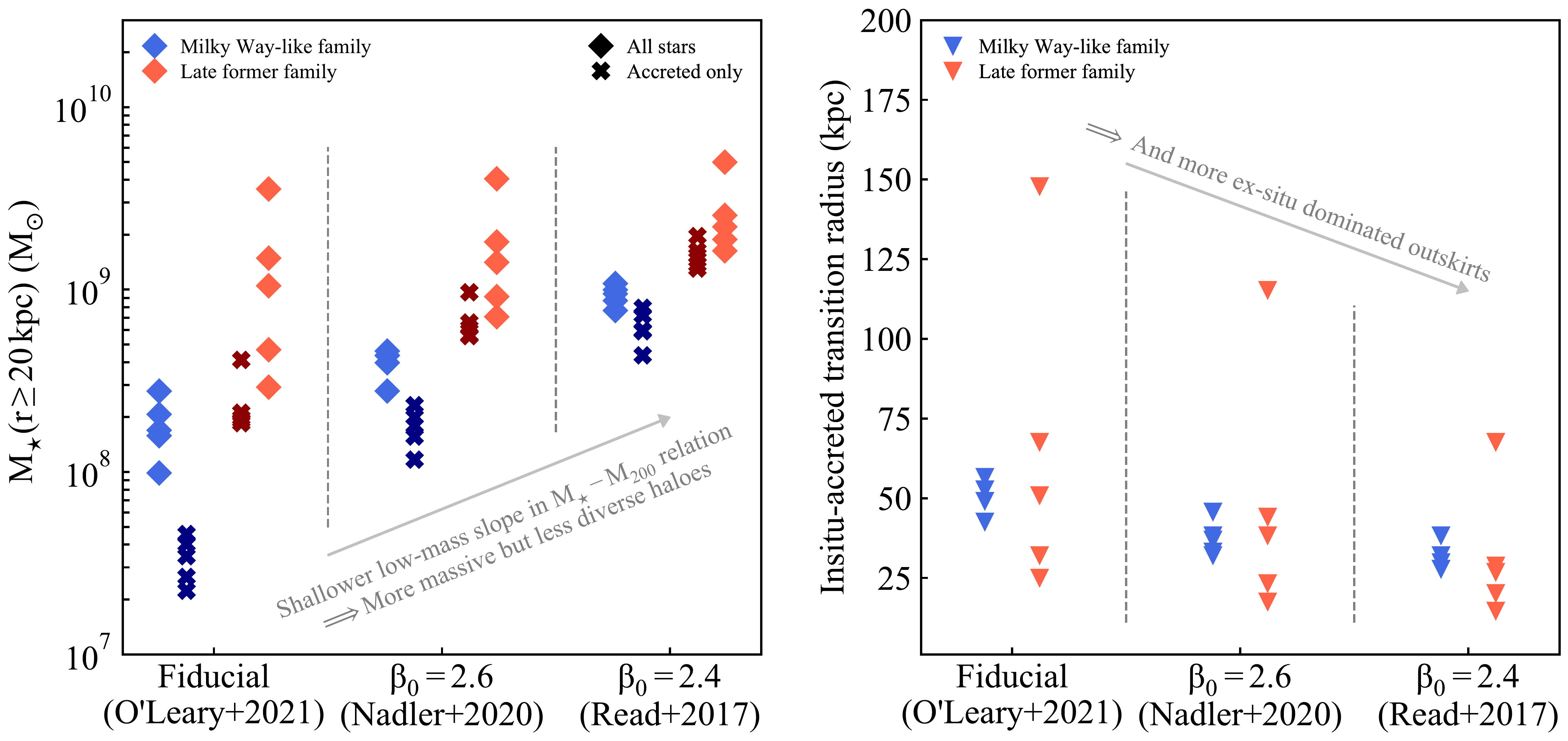}

    \caption{\MR{Impact of varying the dwarf galaxies' stellar-mass-to-halo-mass relation on the stellar halo mass (left) and the  radius at which stellar halos transition from \textit{in-situ} to \textit{ex-situ} dominance (right) for each genetically modified merger scenario (colored symbols).} Increasingly shallower low-mass slopes $\lmslopezero$ (matching \citealt{Nadler2020} and \citealt{Read2017} compared to our fiducial, \MR{left to right in each panel}) increase each infalling dwarf galaxy's stellar mass at fixed halo mass, without affecting the central galaxy. This increase in the accreted mass budget leads to systematically larger stellar halo masses (left), both in the accreted (crosses) and total components (diamonds). Because this mass is primarily deposited in the outskirts, each stellar halo becomes more \textit{ex-situ} dominated, transitioning from \textit{in-situ} to \textit{ex-situ} domination further inwards (right). Reducing the importance of the \textit{in-situ} component in the stellar halo largely diminishes the diversity in observable, total stellar halo masses (vertical spread in left diamonds) from genetically-engineered merger scenarios.
    }
    \label{fig:dwarfsfhsvariations}
\end{figure*}

Turning to the \mwlike family (right-hand panel), we note a different systematic trend: as the last major merger is made larger across family members (from green to purple), the accreted stellar density profiles (dotted) become more centrally concentrated. We quantify this trend by computing the radius at which half of the spherical accreted stellar mass is enclosed (top stars), which is systematically brought inwards as the last major merger becomes larger (grey arrow). Dynamical friction causes larger mergers to sink more efficiently towards the centre and deposit their accreted material further inwards, driving this increase in concentration (\MR{see also e.g. \citealt{White1978, White1980, Barnes1988, Amorisco2017Atlas})}. Consequently, the radial profiles flatten for the histories with a reduced last major merger, yielding a stronger build-up of accreted material in the outskirts \MR{(see also e.g. \citealt{Naab2009, Hopkins2009EllipticalCores, Hilz2013} for similar conclusions in more massive galaxies)}. Since this event is by construction the last significant merger over each galaxy's history, its effect dominates the accreted profile to $z=0$.

In contrast to the accreted component (dotted), the total stellar component (solid) shows reduced variations within this family. The nearly-identical mass growth histories engineered by genetic modifications at late times -- and hence star formation histories through our empirical galaxy formation model (Section~\ref{sec:sec:mergers}) -- combined with the particle tagging technique yield strongly convergent stellar profiles at small radii ($r \lesssim 10 \, \kpc$). Larger differences appear in the stellar halo radial range, though their physical origin is contrasted between scenarios. A larger last major merger (violet and purple) produces a more contracted accreted profile and a more stirred-up \textit{in-situ} component that combine to result in a higher stellar halo mass (Figure~\ref{fig:mshalo}). By contrast, a smaller last major merger (turquoise and green) also produces an increase in stellar halo mass but through a more prominent accreted component in the outskirts, reflected in a flatter accreted profile. These contrasting mechanisms are enough to drive the factor-of-a-few increases in stellar halo fractions compared to the reference \mwlike history (Figure~\ref{fig:mshalo}), but their compensatory nature prevents them from generating the order-of-magnitude diversity that arises when late, violent mergers disrupt an \textit{in-situ} component.

The heating of \textit{in-situ} stars by mergers is thus a key mechanism to generate extended diversity in stellar halo fractions. We stress that, despite dynamical approximations due to the gravity-only context, our particle tagging approach captures the heating and response of an existing \textit{in-situ} component, producing both \textit{in-situ} and \textit{ex-situ} stellar haloes alike those obtained from hydrodynamical simulations (\citealt{LeBret2017,Cooper2017}). The robustness of our findings is nonetheless still sensitive to both (i) the overall amount of external material available to build up the accreted stellar halo, as well as (ii) the dynamical state of the \textit{in-situ} component when mergers infall. Both of these aspects depend on uncertain components of galaxy formation, such as the relationship between stellar and halo masses within dwarf galaxies, and the birth kinematics of stars over a galaxy's history. We now aim to use the flexibility of our approach to efficiently explore how these uncertainties interact with the diversity arising with different mass growth histories. 

\section{The sensitivity of stellar haloes to their dwarf galaxy progenitors} \label{sec:dwarfs}

\begin{figure*}
  \centering
    \includegraphics[width=\textwidth]{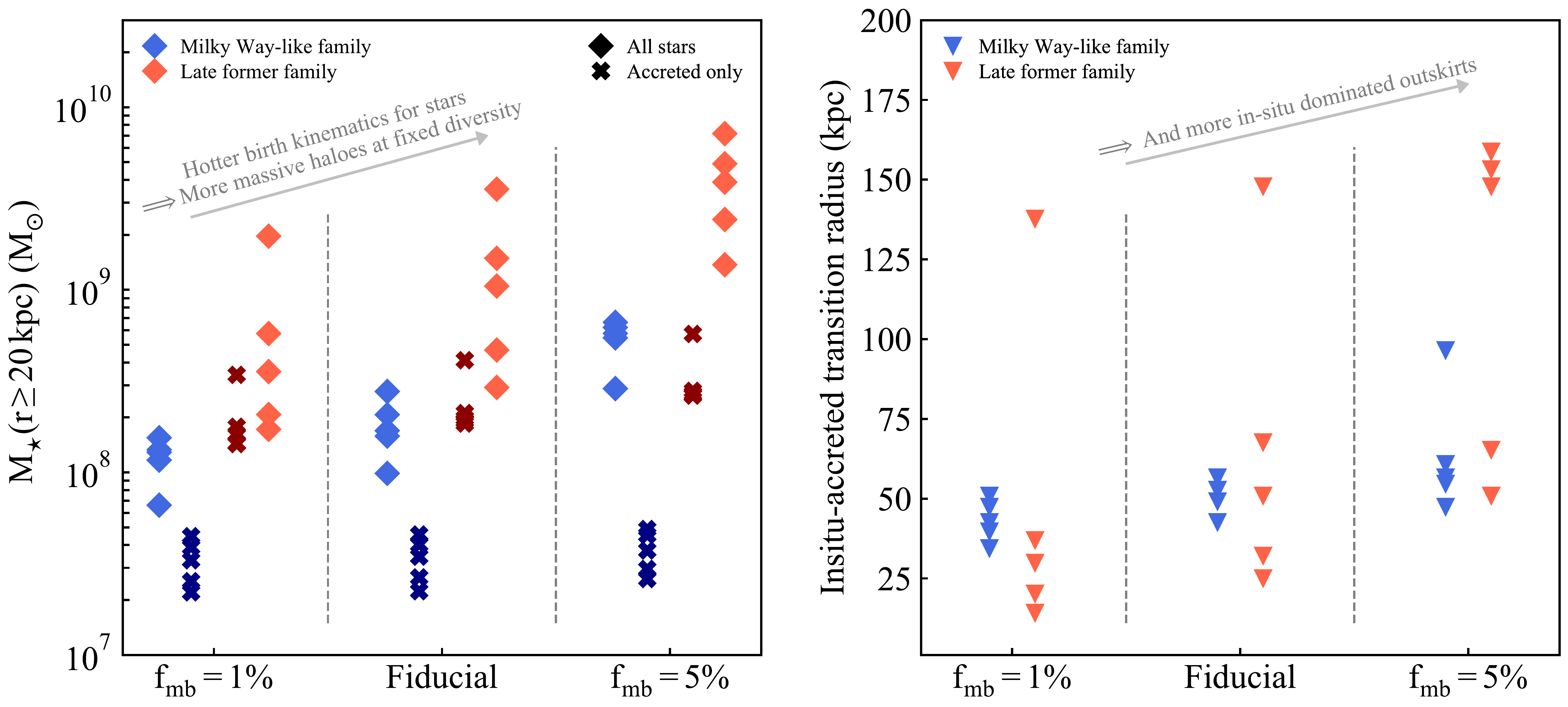}

    \caption{\MR{Impact of varying the birth kinematics of stars on the stellar halo mass (left) and the radius at which stellar halos transition from \textit{in-situ} to \textit{ex-situ} dominance (right).} Allowing newborn stars to form with hotter kinematics (increasing $\fmb$, left to right in each panel) leads to systematically larger total stellar halo masses (left panel, diamonds). Since the accreted stellar halo masses (crosses) are minimally modified, this increase is primarily driven by spreading out the existing \textit{in-situ} component, with each stellar halo transitioning from \textit{in-situ} to \textit{ex-situ} dominance further outwards (right panel). The modified response of \textit{in-situ} stars results in absolute shifts of stellar halo masses, but relative diversity with merger scenarios is conserved (vertical spread in left diamonds).
    }
    \label{fig:fmbvariations}
\end{figure*}

Building blocks of the stellar halo are by nature dwarf galaxies, for which the relationship between stellar masses and halo masses remains debated both in its median and scatter (e.g. \citealt{Brook2014, Read2017, Garrison-Kimmel2017, Munshi2017, Jethwa2018, Nadler2020, Munshi2021}). These uncertainties are reflected in the calibration of our galaxy formation model, and propagate into the amount of accreted material that builds up the stellar halo. In this Section, we quantify how these uncertainties interact with the diversity of possible accretion histories and affect the scatter in stellar halo properties.

To this end, we introduce two alterations from our fiducial star formation histories, specifically targeting dwarf galaxy's stellar masses by varying the low-mass end slope \MR{of the stellar-mass-halo-mass relation} at $z=0$, $\lmslopezero$ (Equation~\ref{eq:lmslope_redshift}), while keeping all other parameters fixed to their best-fitting values. Aiming to bracket an allowed range for $\lmslopezero$, we target two alternate determinations of the stellar mass--halo mass relationship with distinct completeness and environments at the low-mass end, focussing first on Milky Way satellites (\citealt{Nadler2020}) and then on field dwarf galaxies (\citealt{Read2017}). We determine in Appendix~\ref{app:smhm} that shallower $\lmslopezero = 2.6$ and $2.4$ (compared to our fiducial $\lmslopezero = 3.1$) match these respective stellar mass--halo mass relations once star formation histories are recomputed and re-integrated. A shallower low-mass end (higher $\lmslopezero$) yields systematically higher stellar masses for all dwarf progenitors at a given halo mass, with a 2.5-fold (5-fold) increase for $\Mvir = \xMsol{1}{11}$ and 10-fold (130-fold) at $\Mvir = \xMsol{1}{10}$ for $\lmslopezero = 2.6$ ($\lmslopezero = 2.4$ respectively).

\MR{Figure~\ref{fig:dwarfsfhsvariations} shows the response of each stellar halo to these model changes (individual columns in each panel), in their masses (left) and structure by plotting the radius at which halos becomes first \textit{ex-situ} dominated (right)}. As expected, systematically increasing the masses of all progenitors increases every history's stellar halo mass (\MR{left-hand panel}) in both accreted and total components (crosses and diamonds respectively) and across both families of histories (\MR{colored points}). The increase in available accreted material from low-mass satellites primarily builds up the outer light profile, bringing the transition from \textit{in-situ} to \textit{ex-situ} domination inwards (right-hand panel). Shallower stellar mass--halo mass relations thus lead to more massive, and more \textit{ex-situ} dominated stellar haloes.

In addition to the absolute rescaling of masses, the extent of relative scatter between different merger scenarios is reduced for shallower $\lmslopezero$ (vertical spread within each family in Figure~\ref{fig:dwarfsfhsvariations}). This reduction in relative scatter is particularly pronounced for the observable, total component (diamonds), with the range of total stellar halo masses in the \mwlike family (blue) shrinking by a factor 3.0 with increasingly shallower $\lmslopezero$ (going from 0.45 to 0.22 to 0.15 dex for $\lmslopezero = 3.1$, $2.6$ and $2.4$ respectively) and by a factor 2.2 for the \lateformer family (red; going from 1.1 to 0.76 to 0.49 dex). Accreted-only stellar halo masses (crosses) also see their diversity reduced, although more modestly with 20 and 84 per cent diminutions for the \mwlike and \lateformer families respectively.

The physical origin of these reductions is two-fold. (i) Steeper stellar mass--halo mass relations emphasize the importance of massive, rarer events which are the primary targets of our genetic modifications. Increasing the contribution of numerous smaller objects diminishes differences between genetically modified scenarios. Since the total dynamical and stellar mass budget is conserved between objects, this leads to the modest scatter reduction in accreted stellar halo masses. (ii) Total stellar halo masses, however, see their diversity primarily shaped by the varying response of the central component to -- mostly major -- mergers, with the \textit{in-situ} profile dominating the stellar halo mass budget (Figure~\ref{fig:radialprofiles}). As we make stellar haloes more \textit{ex-situ} dominated with shallower $\lmslopezero$, such signatures are lessened, in turn greatly reducing the scatter in observable, total stellar halo masses with varying merger histories.

The sensitivity of stellar halo diversity to the low-mass end slope of the stellar mass--halo mass relation provides an appealing avenue to constrain star formation and feedback processes that shape dwarf galaxy formation. This strongly motivates further explorations with larger samples of merger histories, and beyond the explorative variations in $\lmslopezero$ considered in this work. More physically-motivated alterations should quantify how stellar halo properties are affected by differences in the dwarf population in varying environments (e.g. the impact of the Magellanic Clouds on Milky Way-based estimates of $\lmslopezero$; \citealt{Nadler2020}) and potential correlations between dwarf luminosity functions and assembly histories (\citealt{Lu2016}). We leave this to future studies and focus now on the sensitivity of stellar halo diversity to \textit{in-situ} stars.

\section{The sensitivity of stellar haloes to the \textit{in-situ}, central component} \label{sec:boundfraction}

Our modelling of stellar halo formation relies on the `particle tagging' assumption that stars form tightly bound to their dark matter halo, i.e. in the central region with relatively cold kinematics. This is reflected in the methodology through representing the kinematics and dynamics of newly formed stars by a fixed fraction of the most bound dark matter particles, $\fmb$. The main uncertainty in this process exists of calibrating $\fmb$, with best-fitting values ranging from $\fmb = 1\%$ to $\fmb = 5\%$ depending on which observable or which hydrodynamical modelling is fitted to (\citealt{Cooper2010, Rashkov2012, Cooper2013, LeBret2017, Cooper2017}). Here, we use this freedom to quantify how the birth kinematics of stars affect the diversity of stellar halo properties by systematically varying $\fmb$ while keeping all other model parameters fixed. \MR{Starting from our fiducial $\fmb = 2 \%$, we retag all mass accretion histories with $\fmb = 1 \%$ and $\fmb = 5 \%$ and show in Figure~\ref{fig:fmbvariations} the resulting stellar halo masses and transition radii from \textit{in-situ} to \textit{ex-situ} domination (left- and right-hand panels, respectively) for each family of merger scenarios (coloured points).} 

Accreted-only stellar halo masses (left-hand panel, crosses) are largely conserved following our changes in $\fmb$, with the largest individual change in halo mass reaching 46\% and 17\% in the \lateformer and \mwlike families respectively compared to the fiducial case. These results reaffirm the robustness of accreted density and mass profiles to variations in $\fmb$ (\citealt{Cooper2010, LeBret2017, Cooper2017}).

By contrast, the total stellar halo mass undergoes much stronger variations across our probed range of $\fmb$, with individual masses being modified by up to a factor of 3.7 and 5.6 across the \lateformer and \mwlike families respectively (Figure~\ref{fig:fmbvariations}, left-hand panel, diamonds). For each tagging event, increasing $\fmb$ spreads the same amount of stellar mass over a larger region within each parent dark matter halo. Integrated over the age of the Universe, this effect makes the stellar distribution more extended spatially and less centrally concentrated (see Figure~\ref{fig:resolution_profiles} for a visual example; \citealt{Cooper2013, LeBret2017, Cooper2017}). A larger $\fmb$ therefore leads to a more extended \textit{in-situ} component, pushing the transition from \textit{in-situ} to \textit{ex-situ} domination outwards (Figure~\ref{fig:fmbvariations}, right-hand panel), leading to increasingly massive, and \textit{in-situ} dominated, stellar haloes. 

Despite the large shifts in absolute values described above, the relative diversity of stellar halo masses generated through varying merger histories (left-hand panel, vertical spread within each family) is conserved: for example, going from 0.37 dex with $\fmb = 1 \%$ to 0.44 \MR{with our fiducial model} to 0.36 dex with $\fmb = 5 \%$ in the \mwlike family, and from 1.05 to 1.08 to 0.91 for the \lateformer family. Unlike when increasing the stellar masses of accreted dwarf galaxies through adopting shallower stellar mass--halo mass relation slopes (Section~\ref{sec:dwarfs}), modifying how tightly bound stars are formed within a dark matter halo thus conserves the diversity from different merger scenarios. 

The diversity in stellar halo properties and structures thus responds distinctly to uncertainties associated with the modelling of star formation within the central galaxy on the one hand (this Section), and within its building blocks on the other hand (Section~\ref{sec:dwarfs}). This opens promising avenues to constrain interstellar medium processes at different galactic mass-scales through the population of stellar haloes surveyed by near-future experiments, such as the Nancy Grace Roman Space Telescope, the Vera Rubin Observatory and Euclid.

Our established trends with $\fmb$ further highlight the importance of pinpointing the response of the central galaxy's stellar component to model the absolute values of stellar halo masses. Systematically overmassive and strongly \textit{in-situ} dominated stellar haloes recently observed in cosmological, galaxy formation simulations (e.g. \citealt{Monachesi2019, Font2020, Merritt2020}) could be a sign that \textit{in-situ} stars are not born with sufficiently cold kinematics, either due to the modelling of star formation and feedback within the central host (see also \citealt{Font2020, Bird2021, Keller2021}) or numerical effects affecting galactic kinematics (e.g. \citealt{Ludlow2019DiskHeating, Ludlow2021}).

\section{Conclusion} \label{sec:conclu}

We present and apply a new approach to study how the interplay between cosmological merger histories and galaxy formation physics shapes the build-up and diversity of stellar haloes (Figure~\ref{fig:overview}). We perform dark matter-only, zoomed cosmological simulations of Milky Way-mass hosts and use genetic modifications to create targeted, controlled changes to a host's cosmological mass build-up and merger history (\citealt{Roth2016, Rey2018, Stopyra2021}). To enable wide exploration of merger scenarios, we construct the stellar halo of each dark matter history using (i) post-processed empirical star formation histories based on the \textsc{emerge} model (\citealt{Moster2018, OLeary2021} and (ii) dynamical tracking of stellar kinematics using particle tagging to populate the stellar halo (\citealt{Cooper2010, Cooper2017}).

Targeting two reference Milky Way-mass hosts, we alter their merger histories through four genetically modified initial conditions, with each version reproducing their same reference large-scale, cosmological environment. Each modified galaxy further achieves the same total dynamical and stellar mass as its reference (Figure~\ref{fig:halo715_mah} and Figure~\ref{fig:halo685_mah}), enabling a controlled study where the role of cosmological merger histories in shaping stellar halo structures and observables is cleanly isolated. 

These related scenarios solely varying merger histories achieve an extended scatter in stellar halo mass fractions, reaching 1.5 dex across all objects, comparable to the observed diversity of extragalactic stellar haloes (Figure~\ref{fig:mshalo}). We explicitly show that both earlier and later-forming dark matter histories can lead to similar stellar halo fractions, highlighting their sensitivity to the full history of past progenitors.

Our controlled setup further allows us to demonstrate how increasingly late and violent mergers in galaxy formation histories lead to systematically more massive stellar haloes, dominated by their \textit{in-situ} component further outwards (Figure~\ref{fig:radialprofiles}, left-hand panel). The response of the central, existing \textit{in-situ} stars to different merger scenarios is thus the primary driver of the diversity in stellar halo content. Additionally, more major mergers at early times bring a larger fraction of accreted stars to the inner regions of galaxies, resulting in smaller scatter in stellar halo fractions, as the stellar haloes are predominantly built by subsequent minor events (Figure~\ref{fig:radialprofiles}, right-hand panel). 

The flexibility of our approach also allows for efficient exploration of the sensitivity of these findings to assumptions associated to modelling star formation within both the central galaxy and the building blocks of the stellar halo. First, varying the stellar mass--halo mass relation for dwarf galaxies within present-day constraints, we show how systematically increasing stellar masses in accreted dwarf galaxies leads to more massive, but also more \textit{ex-situ} dominated stellar haloes. This in turn reduces the importance of \textit{in-situ}-driven diversity and thus the scatter in stellar halo masses from varying merger histories (Figure~\ref{fig:dwarfsfhsvariations}). 

Alternatively, allowing stars to form less tightly bound within their dark matter haloes also achieves more massive stellar haloes, through growing larger and more kinematically hot \textit{in-situ} components. By contrast, relative scatter across merger histories is conserved for these variations, with the response of the central galaxy only affecting the absolute stellar halo masses and not their relative spread (Figure~\ref{fig:fmbvariations}).

Our results thus demonstrate that assumptions on how star formation proceeds within the central galaxy on the one hand, and within dwarf progenitor on the other hand, affect distinctly the diversity of stellar halo masses. This strengthens stellar haloes as acute probes of cosmological galaxy formation, and provides exciting prospects to constrain star formation and feedback at different galaxy mass-scales with upcoming integrated light (e.g. \citealt{Rich2019, Danieli2020}) and resolved-stars observations of stellar haloes, for example with the Vera Rubin Observatory, the Nancy Grace Roman Space Telescope, and Euclid. In future work, we plan on building from this work's semi-empirical approach to construct larger samples of cosmological merger scenarios, and extract quantitative constraints on galaxy formation models using mock observations (e.g. \citealt{Sanderson2018, Merritt2020, Sanderson2020, Thomas2021}).

Finally, our findings that more concentrated accreted profiles result from larger last major mergers in a Milky Way-like accretion history could provide insights relevant to ongoing efforts characterising stellar populations and their origin within the central regions of our Galaxy (e.g. in the disc or bulge; \citealt{Helmi2020, Arentsen2020, Lucey2021}). Although accreted profiles are robust predictions of particle tagging approaches at all radii (\citealt{Cooper2010, Cooper2017, LeBret2017}), generalising these to accreted \textit{fractions} in the inner galaxy demands a detailed account of the kinematics of stars within a rotationally supported disc. It requires modelling an expanded set of physical processes to capture secular processes such as the growth of bars and spiral arms (e.g. \citealt{Dobbs2014}) and radial migration through the disc (e.g. \citealt{Sellwood2002, Minchev2010}). In forthcoming work, we will re-simulate our families of initial conditions with state-of-the-art cosmological, galaxy formation zoom simulations (e.g. \citealt{Agertz2020Vintergatan, Renaud2020Vintergatan2, Renaud2020Vintergatan3}), enabling us to quantify how merger histories shape the co-evolution of the inner, central galaxy and its stellar halo.

\section*{Acknowledgements}
We would like to thank \MR{the anonymous referee for a constructive review that improved this manuscript}, Benjamin Moster for sharing details and clarifications on the implementation of the \textsc{emerge} model, and Ethan Nadler and Joseph O'Leary for providing their respective contours in Figure~\ref{fig:sfh_validation}. MR further thanks Oscar Agertz, Lauren Anderson, Ethan Nadler, Andrew Pontzen, Florent Renaud and Yunchong Richie Wang for insightful discussions during the construction of this work and comments on an earlier version of this manuscript. MR is grateful to the CASPEN exchange program that enabled a trip laying down the foundations for this work, and to the Center for Computational Astrophysics at the Flatiron Institute for their hospitality during this visit. MR further acknowledges support from the Knut and Alice Wallenberg Foundation and the Royal Physiographic Society of Lund. The CASPEN Exchange program is partially supported by the UCL Cosmoparticle Initiative and the Center for Computational Astrophysics at the Flatiron Institute. Computations presented in this work were performed on resources provided by the Swedish National Infrastructure for Computing (SNIC) at the Tetralith supercomputer, part of the National Supercomputer Centre, Link\"oping University. The Flatiron Institute is supported by the Simons Foundation. 

\section*{Data availability}

The data underlying this article will be shared on reasonable request to the corresponding author. The IllustrisTNG data used in this article is available at \url{https://www.tng-project.org} and we will share derived data products upon reasonable requests.




\bibliographystyle{mnras}
\bibliography{Stellarhalos} 



\appendix
\section{Star formation histories} \label{app:smhm}

In this work, we assign star formation histories based on the best-fitting, empirical \textsc{emerge} model (\citealt{Moster2018, OLeary2021}, see Section~\ref{sec:sec:emerge}). However, empirical models are calibrated by linking a statistical population of dark matter haloes to the observed properties of the galaxy population, an unfeasible procedure to apply to our handful of genetically modified, high-resolution, zoomed cosmological simulations. In this Appendix, we demonstrate that fixing the \textsc{emerge} parameters to their published best-fitting values (Table~\ref{table:bestfitparams}, \citealt{OLeary2021}) and applying them without recalibration to this different context produces consistent star formation histories and stellar masses for all galaxy progenitors. This reaffirms a similar approach independently developed \MR{by} \citealt{Wang2021}, \MR{who} applies the best-fitting parameters of the \textsc{universemachine} empirical model to deep dark matter-only, zoomed simulations. \MR{In addition to using the best-fitting parameters, we} use the analytical nature of empirical star formation histories to introduce parameter variations \MR{and thus explore} remaining uncertainties in our modelling.

\begin{figure}
  \centering
    \includegraphics[width=\columnwidth]{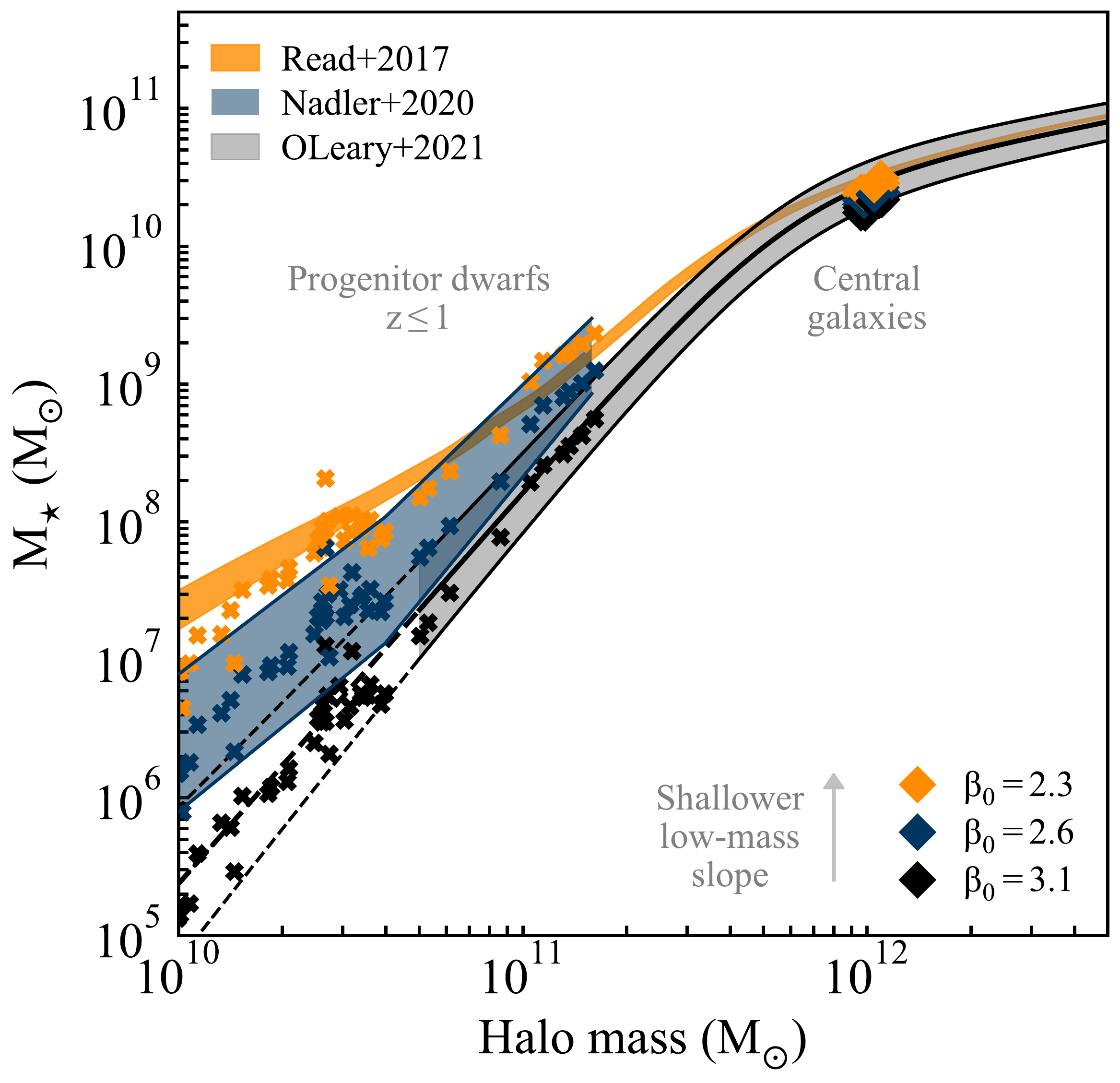}

    \caption{Stellar masses as a function of peak halo masses for the central galaxies presented in this work (diamonds) and dwarf galaxies that infall within a host after $z=1$ (crosses). Extrapolating the \textsc{emerge} best-fitting parameters to individual, zoomed simulations (black symbols) produces stellar masses consistent with the one-sigma contour of the \textsc{emerge} population (grey contours). A shallower low-mass end slope of the baryon conversion efficiency, $\lmslopezero$, increases the stellar masses of each progenitor dwarf at fixed halo masses (blue and orange crosses), matching alternative literature determinations of this mapping (blue and orange contours). This allows us to test how uncertainties in $\lmslopezero$ propagate onto stellar halo observables (Figure~\ref{fig:dwarfsfhsvariations}).
    }
    \label{fig:sfh_validation}
\end{figure}

Figure~\ref{fig:sfh_validation} shows the relationship between stellar masses and halo masses for all galaxies presented in this work. We extract central galaxies (diamonds), together with each of their merger tree progenitors (i.e. dwarf galaxies that merge with the central, main progenitor; crosses) after $z=1$ for which we show $\Mvir$ at first infall. We compare these individual galaxies with the integrated stellar mass--halo mass relation at $z=0$ resulting from \textsc{emerge} parameters as in Table~\ref{table:bestfitparams} (J. O'Leary, private communication; see \citealt{Moster2018}, table 8 for previous parameter iteration). All stellar masses are consistent with the median and one-sigma contour determined from fitting to a large statistical population (black contour), \MR{confirming} that our alterations to the \textsc{emerge} prescriptions and context produces results compatible with the best-fitting model. 

This verification, however, only ensures that our results are consistent within the \textsc{emerge} framework. Building blocks of the stellar halo are by nature dwarf galaxies, which are at the frontier of observational data used to calibrate the model (dashed black lines show extrapolations beyond the calibration range). In fact, significant uncertainties remain in the relationship between stellar mass and halo mass at the low-mass end, both in the median, scatter of the relation, and in potential environment or host dependencies (e.g. \citealt{Brook2014, Read2017, Garrison-Kimmel2017, Munshi2017, Nadler2020, Munshi2021}). Furthermore, empirical models might not be flexible enough in their parametrization to capture relevant quenching mechanisms in the smallest dwarf galaxies considered in this work (\citealt{Wang2021}).

\begin{table}
  \centering
  \caption{Parameters used in this work to assign empirical star formation histories to dark matter haloes. Our fiducial model (second column) has identical values to the best fit obtained by \citet{OLeary2021}, who compare a large halo sample to observables of the galaxy population. We explore remaining uncertainties on dwarf galaxy scales by introducing two variations of the low-mass end slope $\lmslopezero$ (third and fourth columns) engineered to match stellar mass--halo mass relations from \citet{Nadler2020} and \citet{Read2017} respectively (Figure~\ref{fig:sfh_validation}).}
  \label{table:bestfitparams}

  \begin{tabular}{l c c c}
    \hline
    SFH parameter & Fiducial & Nadler+2020 & Read+2017 \\
    \hline

    $\Mzero$ & $11.34829$ & = & = \\

    $\Mz$ & $0.654238$ & = & = \\

    $\effzero$ & $0.009010$ & = & = \\

    $\effz$ & $0.596666$ & = & = \\

    $\lmslopezero$ & $3.094621$ & $2.64$ & $2.34$ \\

    $\lmslopez$ & $-2.019841$ & = & = \\

    $\hmslopezero$ & $1.107304$ & = & =\\

    $\hmslopez$ & $0.$ & = & = \\
    
    \hline
  \end{tabular}
\end{table}

We illustrate these uncertainties in Figure~\ref{fig:sfh_validation}, showing two alternative data-driven determinations of the stellar mass--halo mass relation, specifically focussed on isolated dwarf galaxies (\citealt{Read2017}, orange\footnote{Shading is defined between the `high' and `low' determinations in \citet{Read2017} and does not represent a confidence interval or population scatter.}) and satellites of the Milky Way (\citealt{Nadler2020}, blue). These two relations and our fiducial model bracket a factor of 2 interval in stellar masses for dwarfs with $\Mvir \approx \xMsol{1}{11}$, extending to two order of magnitudes discrepancies at the lowest dynamical masses. 

To quantify how these uncertainties governing crucial elements of modelling galaxy formation propagate to the building of stellar haloes, we introduce a systematic variation of the normalisation of the low-mass end slope at $z=0$, $\lmslopezero$, keeping all other parameters fixed. In this way, we modify the star formation histories within dwarf galaxies with minimal alterations to the central's stellar mass build up. We select two variations, $\lmslopezero = 2.6$ (blue) and $\lmslopezero = 2.4$ (orange) from our fiducial $\lmslopezero = 3.1$, obtained respectively by extracting the slope from the median stellar mass--halo mass relation in \citet{Nadler2020}, and iterating from a visual comparison between the resulting stellar masses and infall halo masses in Figure~\ref{fig:sfh_validation} and the stellar mass--halo mass relation in \citealt{Read2017}. 

As we make $\lmslopezero$ shallower (black to blue to orange), each progenitor dwarf's stellar mass increases at fixed halo mass, bringing it in broad agreement with its respective stellar mass--halo mass determination. Since we only modify the low-mass end behaviour of our star formation histories, stellar masses of all central galaxies vary at most by 15 per cent at fixed halo mass. We stress that our approach should not be interpreted as a quantitative, physically-motivated comparison between models of dwarf galaxy formation, but rather as an efficient and simple way to test the importance of dwarf galaxy stellar masses in building the \textit{ex-situ} stellar halo at fixed \textit{in-situ} contribution (Section~\ref{sec:dwarfs}). 

\section{Sensitivity to numerical effects} \label{app:numerics}

\begin{figure*}
  \centering
  \includegraphics[width=\textwidth]{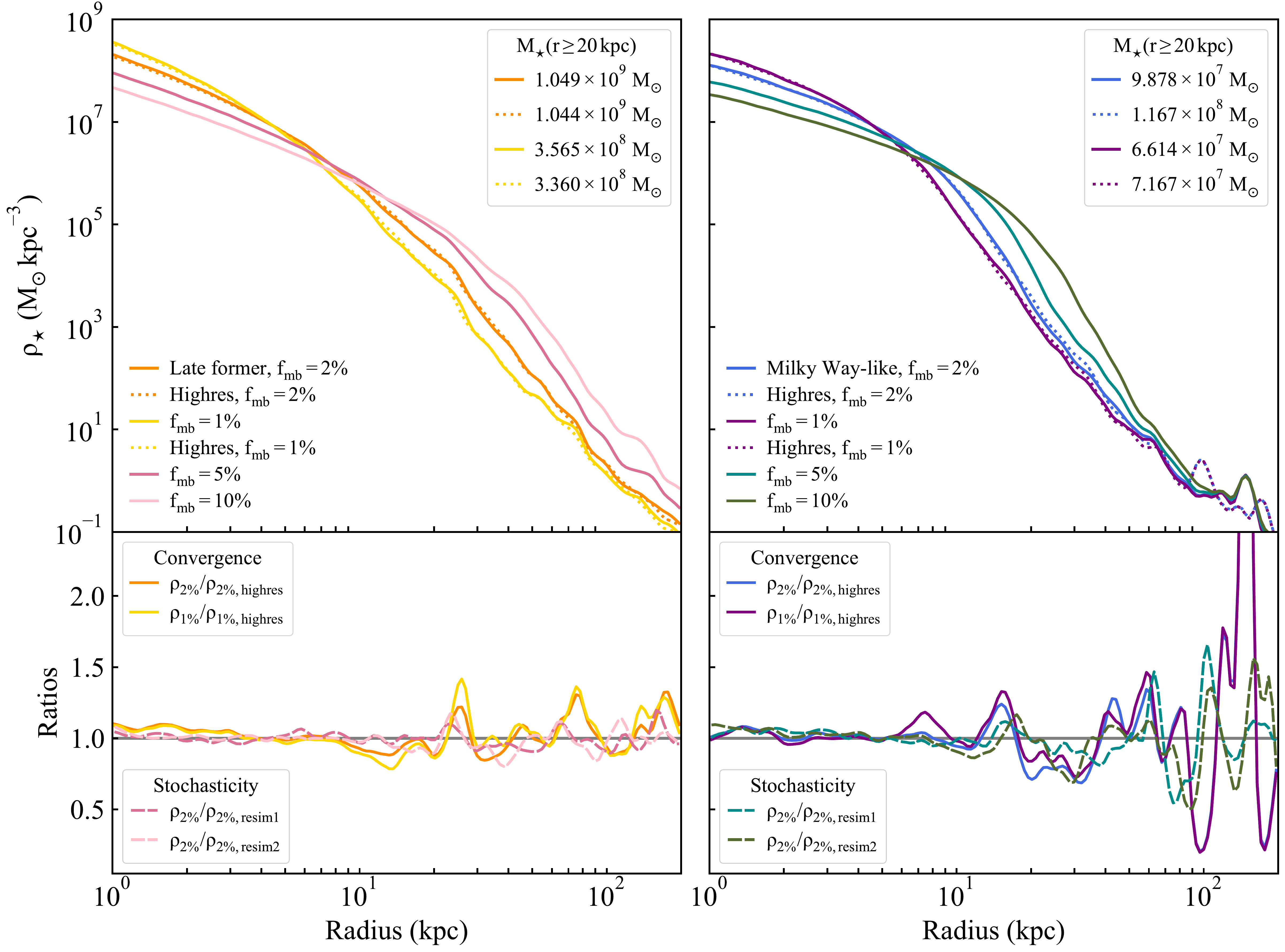}
  \caption{Stellar density profiles at fiducial (solid) and increased (dashed) numerical resolutions for the reference histories in the \lateformer (left) and \mwlike (right) families \MR{when varying resolution and the fraction of most bound particles}. Both profiles are converged within 50 per cent up to $r\sim 80 \, \kpc$ (bottom panels), leading to robust integrated stellar halo masses (top right boxes). Discrepancies at larger radii arise from slightly different orbits and final positions of bound dwarf galaxies within the virial volume, and do not affect this robustness. Both resolution and run-to-run stochasticity (bottom panels, dashed) are subdominant sources of scatter compared to merger histories.
  }
 \label{fig:resolution_profiles}
\end{figure*}

In this Appendix we quantify the sensitivity of our genetically-modified stellar haloes to two numerical factors, resolution and chaotic amplification of truncation errors.  

Focusing first on numerical resolution, we re-simulate the reference mass accretion histories in both families of initial conditions with 8 times more dark matter particles ($\mdm = \xMsol{2.9}{4}$ compared to $\xMsol{2.3}{5}$), allowing refinement down to $\softening = 35 \, \pc$ (compared to $\softening = 140 \, \pc$). Star formation and particle tagging histories are produced identically to our fiducial model (Section~\ref{sec:methods}) to obtain stellar density profiles at $z=0$ which we compare in Figure~\ref{fig:resolution_profiles} (fiducial and higher resolution in solid and dotted respectively).

Within $r \lessapprox 80 \, \kpc$, the higher-resolution stellar density profile of each object is within 50 per cent of its fiducial counterpart (ratios in bottom panel). This inner region dominates the stellar halo mass budget, leading to total stellar halo masses differing by 18 per cent in the \mwlike history and 0.4 per cent in the \lateformer history (top-right boxes). To illustrate the good agreement across both numerical resolutions, we contrast these variations in each panel with those, much larger, introduced by our exploration of allowed parameter space for the fraction of most bound dark matter particles, $\fmb$ (solid curves; see also Section~\ref{sec:boundfraction} and Figure~\ref{fig:fmbvariations}). Systematically varying $\fmb$ at fixed resolution yield order-of-magnitude changes in stellar profiles at a given radius, in turn leading to factor-of-a-few modifications to the total stellar halo masses (top right box). This conclusion further holds for two distinct $\fmb$ ($1\%$ and $2\%$ in Figure~\ref{fig:resolution_profiles}) at both resolutions, excluding any compensation effects due a potential coupling between resolution and $\fmb$.

In contrast with the established convergence in the inner radial range, stronger mismatches in the stellar density profiles are apparent at $r \gtrapprox 80 \, \kpc$. These discrepancies arise from small variations in positions of bound dwarf galaxies at $z=0$ between the two resolutions -- they are visible in both reference objects, but stand out particularly for fainter stellar haloes (right-hand panel) where the background, smooth, component drops more rapidly with radius, thus emphasizing the importance of bound peaks in the outskirts. Achieving convergence in the orbital evolution of the smallest dark matter haloes remains a significant challenge for cosmological simulations, due to the propagation of numerical instabilities when following infall and stripping with an adaptive force resolution (\citealt{vandenBosch2018}). Since our main observable, the integrated stellar halo mass, is largely unaffected, we conclude that numerical resolution does not affect our results, but we caution that further studies targeting the satellite population will require a detailed quantification of this source of uncertainty. 

Due to the chaotic and non-linear nature of gravitational collapse, stochastic variations from run to run in numerical simulations of structure formation are expected (\citealt{Genel2019, Keller2019}). Since an increase of resolution also seeds new chaotic perturbations, we now disentangle their magnitude alone and verify that our conclusions are unaffected by this source of variance. We re-simulate the reference mass growth histories from their initial conditions two additional times with the exact same numerical setup as described in Section~\ref{sec:sec:suite}. Our simulation code \textsc{ramses} does not conserve the arithmetic order of operations from run to run, seeding distinct round-off errors in each simulation instance and hence chaotic perturbations propagating over the course of a simulation. 

The bottom panel of Figure~\ref{fig:resolution_profiles} shows the stellar density profiles of each re-simulation, divided by their first reference. Stochastic variations from run to run introduce noise in the profiles, within 50 per cent of the reference run at all radii for each instance. These variations are comparable or smaller than those introduced by our increased numerical resolution, in which both sources of error are tested at once. We further verified that stellar halo masses are converged within 10 per cent between all re-simulations. We therefore conclude that numerical stochasticity and resolution are subdominant sources of variance in our stellar halo masses, compared to physical factors considered in this work such as varying merger histories (Section~\ref{sec:stellarhaloes}) or star formation modelling (Section~\ref{sec:dwarfs} and ~\ref{sec:boundfraction}).

\section{A suite of randomly-selected Milky Way mass haloes} \label{app:randomsuite}

\begin{figure}
  \centering
  \includegraphics[width=\columnwidth]{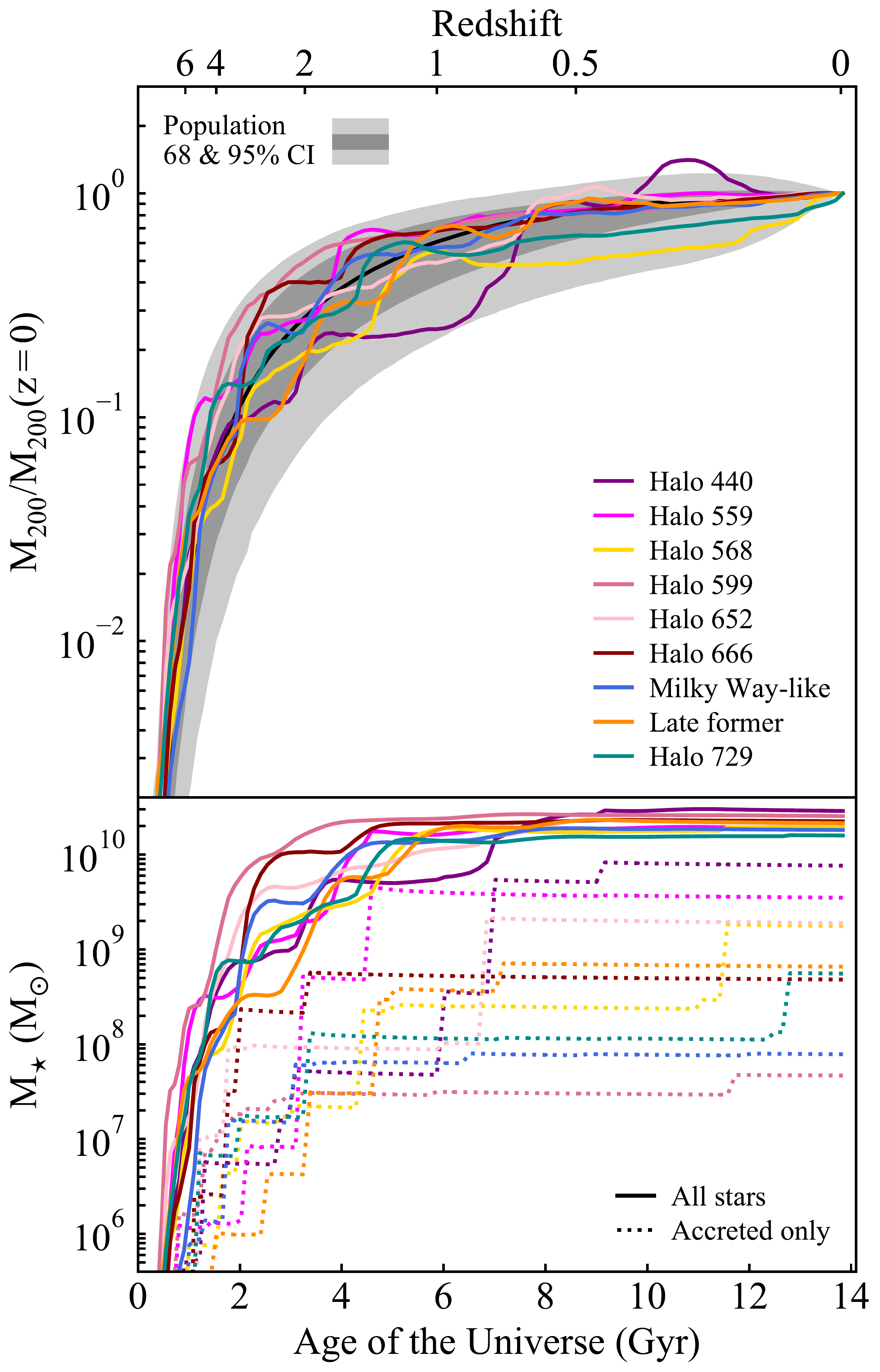}
  \caption{\MR{Fractional growth of dynamical mass (top panel) and total and accreted stellar mass (bottom panel, plain and dotted respectively) over cosmic time, across a suite of nine randomly-selected Milky Way-mass haloes ($\xScientific{8}{11} \leq \Mvir \leq \xMsol{2}{12} $). Haloes used as reference for our genetic modifications (blue and orange, Section~\ref{sec:gms}) are within the scatter of individual mass growths across the suite, thus having typical assemblies in both dynamical and stellar mass.}
  }
 \label{fig:suitemahs}
\end{figure}

\MR{
In this Appendix, we present nine isolated, randomly-selected Milky Way-mass haloes from which we select the reference haloes of our genetically-modified families in Section~\ref{sec:sec:suite}. All objects are evolved using the same numerical setup and resolution as in Section~\ref{sec:sec:suite}, and their stellar haloes are constructed using the procedure described in Section~\ref{sec:sec:emerge} and~\ref{sec:sec:cooper}. We further use this suite of independent object to quantify the robustness of our interpretation and results against a small population of simulated stellar halos generated with the same approach. Furthermore, Appendix~\ref{app:tng} discusses a comparison against a larger sample, but with different resolution and physical model which inhibits direct comparisons.

Figure~\ref{fig:suitemahs} shows the fractional dynamical mass growth histories of our nine galaxies (top panel), and their build-up of stellar mass over cosmic time (bottom panel). All objects within the suite are inside the 95\% confidence interval of the large sample of mass growth histories from TNG-300-1-DMO (grey contour, see Section~\ref{sec:sec:suite}). The reference haloes used as starting points for our genetic modifications (blue and orange) are within the random scatter of stellar and dynamical mass growth histories across the suite. This validates that our reference histories are fairly typical realizations of a $\Lambda$CDM Universe.

Furthermore, we can test the physical insights gained through our genetically-modified experiments (Section~\ref{sec:stellarhaloes}) against these independent merger histories. In particular, Halo 568 (yellow in Figure~\ref{fig:suitemahs}) is another Milky Way-like history, with a major merger at $z\approx2$ and mass ratio of 1:3, and a quiet assembly afterwards. This mass ratio puts it between the reference halo of the \mwlike family (mass ratio 1:4) and its modified “GM:Larger last major merger” counterpart (mass ratio 1:2). When constructing its stellar halo, we obtain a stellar halo mass fraction of 0.0098 (for $\Mstar = \xMsol{2}{10}$), i.e. in between the reference \mwlike and its GM (mass fractions of 0.0054 and 0.011 respectively). Moreover, our latest forming halo (Halo 440, purple in Figure~\ref{fig:suitemahs}) that undergoes very late major mergers is also the largest stellar halo in our sample, with a mass fraction of 0.17, in line with our results obtained for the \lateformer family that later major mergers yield more massive stellar haloes. These \textit{a-posteriori} verifications on independent objects bring further robustness to the trends we find between stellar halo masses, structures and merger histories described in Sections~\ref{sec:sec:smasses} and~\ref{sec:sec:radialstructure}.
}

\section{Comparison with a larger sample of simulated stellar haloes} \label{app:tng}

\begin{figure}
  \centering
  \includegraphics[width=\columnwidth]{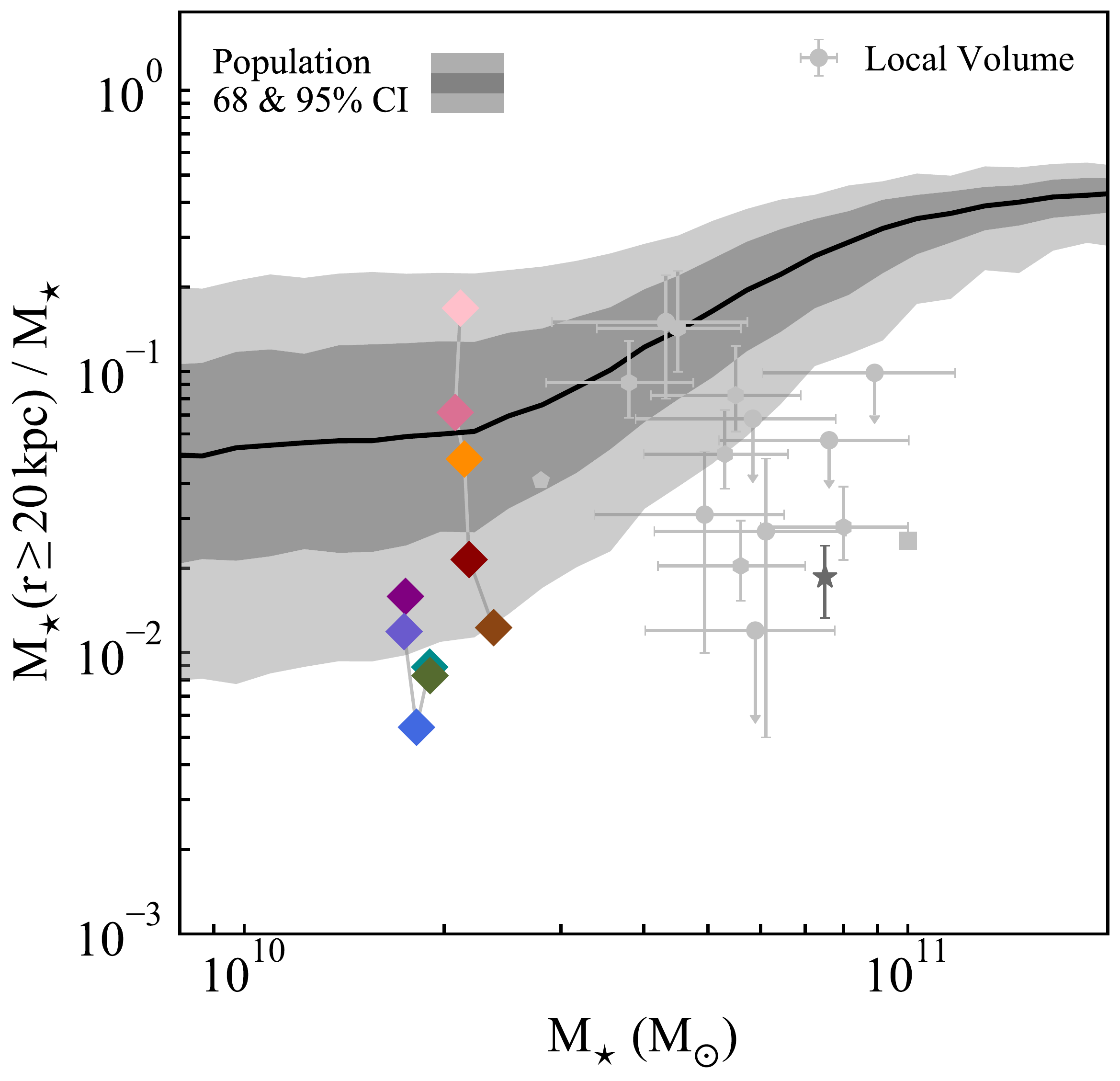}
  \caption{\MR{Same as Figure~\ref{fig:mshalo}, but contrasting our genetically-modified scenarios (colored points) against a large sample of galaxies obtained from Illustris-TNG300 (grey contours).}
  }
 \label{fig:tngpopulation}
\end{figure}

\MR{In Section~\ref{sec:sec:smasses}, we construct qualitative estimates of the significance of the diversity obtained across our genetically modified scenarios. We base these estimates on a small, controlled sample of independent simulated objects using the same exact approach as our \lateformer and \mwlike families (Appendix~\ref{app:randomsuite}), and a handful of observed stellar halos in the Local Volume (Figure~\ref{fig:mshalo}). In this Appendix, we further compare the magnitude of our variations in stellar halo fractions against a large population of simulated stellar haloes using an independent approach. 

We use the hydrodynamical counterpart to the dark matter-only population used in Section~\ref{sec:gms}, namely `TNG-300-1', for which we extract pre-computed stellar density profiles from the analysis described in \citet{Diemer2018, Diemer2019}. We then compute the mass outside 20 kpc to obtain stellar halo fractions as in Section~\ref{sec:sec:smasses} and show in Figure~\ref{fig:tngpopulation} the median and 68--95\% confidence intervals in 40 bins of stellar masses equally spaced in log (grey contours). 

As when compared to the observational scatter (grey points, same sample and coding as Figure~\ref{fig:mshalo}) and our control suite of randomly selected haloes, the \lateformer family extends across a large fraction of the population scatter, reaching from the lower to the upper 2$\sigma$ contour of the TNG population. This emphasizes the extent of scatter in stellar halo masses obtained when varying the timing of late, major mergers.  

The \mwlike family shows reduced scatter, within the lower 2$\sigma$ contour or rarer for their stellar masses. While these objects appear as outliers of the TNG population, we stress that, despite the inclusion of hydrodynamics, this simulation is known to overproduce stellar mass in the outskirts of Milky Way-mass galaxies compared to observations (\citealt{Merritt2020}, see also e.g. \citealt{Monachesi2019} for similar findings with a different galaxy formation model). Such mismatches motivate a more detailed understanding of how model parameters map onto the diversity of stellar haloes, which is the focus of Sections~\ref{sec:dwarfs} and~\ref{sec:boundfraction}.}

\bsp	
\label{lastpage}
\end{document}